\documentclass[letterpaper,dvipsnames]{article}
\pdfoutput=1
\usepackage{jheppub}
\usepackage{amsmath}
\usepackage{tensor}
\usepackage{graphicx}
\usepackage{array}
\usepackage{ulem}
\usepackage{cancel}
\usepackage{xcolor}
\usepackage{slashed}
\usepackage{afterpage}
\usepackage{appendix}
\usepackage{multirow}
\usepackage{float}
\usepackage{mathtools}
\mathtoolsset{centercolon}

\usepackage{multirow}

\usepackage{comment}
\usepackage{multirow}

\newcommand{\ads}{{\rm{AdS}_4}}

\newcommand{\uv}{{\rm uv}}
\newcommand{\ir}{{\rm ir}}

\newcommand{\eff}{{\rm eff}}

\newcommand{\bulk}{{\rm bulk}}

\newcommand{\Fp}{F_{\Phi_+}}
\newcommand{\Fm}{F_{\Phi_-}}
\newcommand{\Fpm}{F_{\Phi_\pm}}
\newcommand{\Fcp}{F_{C_+}}
\newcommand{\Fcm}{F_{C_-}}
\newcommand{\Fcpm}{F_{C_\pm}}
\newcommand{\bFp}{F^\dag_{\Phi_+}}
\newcommand{\bFm}{F^\dag_{\Phi_-}}

\newcommand{\bFcp}{F^\dag_{C_+}}
\newcommand{\bFcm}{F^\dag_{C_-}}

\newcommand{\Fcf}{F_{C_4}}
\newcommand{\C}{\mathbf{C}}

\newcommand{\dsq}{\int d^2 \theta \,}
\newcommand{\dfour}{\int d^4 \theta \,}
\newcommand{\MP}{M_p}

\title{Sequestered Conformal Anomaly Mediation (SCAM)}

\author{Michael Nee,}
\author{Lisa Randall}

\affiliation{Department of Physics, Harvard University, Cambridge, MA, 02138, USA}
\emailAdd{mnee@fas.harvard.edu}
\emailAdd{randall@g.harvard.edu}

\abstract{
Supersymmetric models in singular extra dimensional spaces feature prominently in many interesting phenomenological models, including those derived from string theory. In this paper we explicitly derive the low energy theory of phenomenologically viable supersymmetric theories in five dimensions, and highlight several aspects of these models that are not obvious from working solely in the 4d effective theory. Important deviations arise for anomaly mediation, which is purported to be a predictive mechanism to mediate supersymmetry (SUSY) breaking that is naturally most relevant in extra dimensional theories. Despite this, most analyses of anomaly mediation have been performed in the 4d effective theory. We fill this gap in the literature by constructing stabilized supersymmetric theories in 5d. Studying Sequestered Conformal Anomaly-Mediated (SCAM) in full generality reveals important  deviations from the 4d EFT expectations, particularly for the role of boundary superpotentials, the radion and the predicted universality of anomaly mediation. We discuss the requirements for viable extra dimensional models of SUSY-breaking, and demonstrate when and how the anomaly-mediated masses in 5d reduce to the na\"ive 4d supersymmetric result. In many cases supersymmetry is necessarily broken at the 5d level, leading to anomaly-mediated and other supersymmetry breaking masses that are not derivable in a simple supersymmetric 4d EFT, but need to be included as matching corrections. We comment on the potential implications of our methods for phenomenology and singular higher-dimensional constructions, such as the KKLT scenario.}

\begin{document}

\maketitle

\section{Introduction}

The study of supersymmetry (SUSY) on singular spaces in higher dimensions features prominently  in several contexts. All phenomenological models coming from string theory involve extra dimensions and supersymmetry. These models often involve singular compactifications, with examples being orientifold projections in type II constructions~\cite{Dai:1989ua, Horava:1989vt}, orbifold compactifications of heterotic models~\cite{Dixon:1985jw, Dixon:1986jc}, and the strongly coupled limit of the heterotic theory~\cite{Horava:1995qa, Horava:1996ma}. A widely studied example is the KKLT scenario~\cite{Kachru:2003aw} -- designed to address the question of stabilization and the small cosmological constant -- which involves a gaugino condensate on a singularity.

Furthermore, extra-dimensional models with branes have many attractive phenomenological features. Warped geometries offer a way to address the hierarchy problem~\cite{Randall:1999ee, Randall:1999vf} and supersymmetric extra dimensions can address some problems in unified theories~\cite{Kawamura:1999nj,Kawamura:2000ev,Hall:2001pg,Hall:2001xb,Hall:2002ci,Hebecker:2001wq,Hebecker:2001jb,Altarelli:2001qj}. Higher-dimensional anomaly-mediated models employ branes localized in extra dimensions to separate the Standard Model~(SM) from the SUSY-breaking sector, thereby addressing some issues of flavor while communicating SUSY breaking to the supersymmetric Standard Model (MSSM) fields~\cite{Randall:1998uk, Giudice:1998xp}. Anomaly mediation is of particular interest in singular higher-dimensional spaces, where it can be the leading supersymmetry-breaking communication. It is also important because it gives rise to superpartner masses that depend primarily on gauge charges, avoiding dangerous flavor-violating interactions, and because it allows for a non-vanishing gaugino mass even in the absence of singlets.

Because anomaly-mediated mass terms are suppressed both by beta functions and inverse powers of the Planck scale, $\MP$, they are typically smaller than masses from contact terms. Anomaly-mediated SUSY-breaking (AMSB) is therefore most relevant when the theory has a `sequestered' form, in which contact terms between the MSSM and the hidden sector (including higher-dimensional operators) are suppressed. While this generically requires fine-tuning, it is a natural expectation in extra-dimensional models in which the two sectors are physically separated in the higher-dimensional theory~\cite{Randall:1998uk}. Anomaly mediation can also be relevant in purely 4d theories with conformal sequestering~\cite{Luty:2001jh,Luty:2001zv,Sundrum:2004un, Schmaltz:2006qs}, but when true we expect a corresponding 5d holographic interpretation so that the results should be a subset of what we do below.

It has been argued that when anomaly mediation is the dominant source communicating supersymmetry-breaking the 4d theory is highly predictive, as all the SUSY-breaking terms are calculable in terms of IR quantities up to a single unknown scale -- the $F$-term of the conformal compensator~($F_C$). Phenomenological string models also often contain the features required for anomaly mediation to be important, providing a top-down motivation for understanding the details of AMSB as well~\cite{Kachru:2007xp, Berg:2010ha}.

Because of its significance and  interesting theoretical features, a significant amount of work has been devoted to determining the phenomenology of anomaly mediation~\cite{Pomarol:1999ie, Chacko:1999am, Katz:1999uw, Bagger:1999rd, Luty:1999cz, Bagger:2000dh,Luty:2000ec, Allanach:2000gu, Carena:2000ad, Gherghetta:2000qt, Chacko:2000fn, Kaplan:2000jz, Gherghetta:2000kr, Arkani-Hamed:2000znc, Luty:2001jh, Luty:2001zv, Chacko:2001jt, Nelson:2001ji, Luty:2002ff, Luty:2002hj, Rattazzi:2003rj, Sundrum:2004un, Arkani-Hamed:2004zhs, Gregoire:2004nn,Choi:2005uz, Choi:2005ge, Ibe:2005pj, Ibe:2005qv, Schmaltz:2006qs, Kachru:2007xp,Dine:2007me, Son:2008mk, Gripaios:2008rg, Rattazzi:2009ux, Berg:2010ha, DEramo:2012vvz, DEramo:2013dzi, Heidenreich:2014jpa, Baer:2018hwa, Baer:2025zqt}. Many models draw inspiration from 5d theories with warped extra dimensions, or explicit string-theoretic models such as KKLT (and related models)~\cite{Klebanov:1998hh, Verlinde:1999fy, Klebanov:2000nc, Klebanov:2000hb, Kachru:2003aw, Kachru:2019dvo} or M-theory~\cite{Horava:1995qa, Horava:1996ma, Horava:1996vs, Lalak:1997zu, Nilles:1997cm,Antoniadis:1997xk, Lukas:1998yy, Lukas:1998tt, Lukas:1998ew, Lukas:1999kt, PaccettiCorreia:2006rpo, PaccettiCorreia:2007ret}. Despite this, almost all the work on anomaly mediation has been performed in the 4-dimensional effective theory. In this work we conduct a study of anomaly mediation in a complete 5d theory, without resorting to a 4d EFT analysis. The closest paper to our work is ref.~\cite{Son:2008mk} which set up the model for anomaly mediation in a 5d theory (though they ultimately calculated anomaly-mediated masses in 4d). Although we agree in principle with their approach we disagree with some aspects of their 5d analysis,
including an inconsistent gauge choice, treating the compensator as a dynamical field, and not including the full solutions to the 5d equations of motion. 

The motivation for our work stems from several puzzles and discrepancies in the literature about anomaly mediation when derived from extra dimensions. First, one might question whether the four-dimensional effective theory answer for anomaly mediation always correct. In four dimensions, anomaly mediation relies on a constant superpotential that sources $F_C$ and generates the negative energy required for a supersymmetric AdS space. In 5d in principle there can be independent bulk and boundary superpotentials, and it is not clear how $F_C$ relates in either the 5d or the 4d theory.\footnote{In principle one can derive anomaly-mediated masses without the conformal compensator formalism, but this has been done only for four-dimensional theories.} In higher dimensions, the 4d cosmological constant must be the same at every point in the extra dimension,\footnote{We discuss the implications of this fact in a non-supersymmetric theory in a companion paper~\cite{Lust:2025vyz}.} in which case it is also not clear how to consistently generate a cosmological constant throughout the bulk. 
It is therefore not obvious how an $F_C$ in 5d, sourced by bulk and boundary superpotentials, relates to a constant single $F_C$ in 4d. 
Some papers argue for non-universal masses from boundary superpotentials~\cite{Luty:1999cz, Luty:2000ec}, but did not derive the full corresponding 4d EFT, which we will see is highly model-dependent. 

Other questions revolve around the role of the radion and whether the radion contribution to SUSY-breaking terms can be isolated from the compensator contribution. Furthermore, brane-localized superpotentials are associated with singular points in the extra dimension and it is not clear how these singularities are reflected in the 4d potentials. 

For all these reasons, a more complete exploration of supersymmetry breaking on 5d orbifolds is warranted. In this paper we perform the study of anomaly mediation by doing the full analysis in 5d AdS space, the supersymmetric generalization of the Randall-Sundrum (RS) scenario~\cite{Randall:1999ee, Randall:1999vf}. Our work addresses supersymmetry breaking, stabilization, and anomaly mediation in the context of a single warped extra dimension -- with occasional comments on flat extra dimensions. We address how supersymmetry is implemented, how supersymmetry breaking can be communicated, and how to generate the necessary negative energy to allow for flat space (in the 4d theory) after SUSY-breaking. 
\\

A critical feature of the 5d theory is that the K\"ahler potential for the radion, $\Sigma$, has a no-scale form~\cite{Cremmer:1983bf, Ellis:1984kd, Lahanas:1986uc}. The $F_\Sigma$ equation of motion then relates $F_C$ to products of the hypermultiplet $F$-terms times scalar field profiles. If the model is truly no-scale there is no stabilization mechanism, meaning that both $F_C$ and the potential vanish. 
However, loop corrections to the cosmological constant will lead to a positive energy density after SUSY is broken.
Even when the fields are nonzero, $F_C$ is suppressed (by field values in units of the 5d Planck scale, $M_5$) relative to SUSY-breaking auxiliary fields. The no-scale structure must therefore be broken in order to stabilize the model, generate an AdS$_4$ solution, and a nonzero $F_C$ that can generate anomaly-mediated masses. It is therefore of particular significance to our results how this no-scale structure is broken.

There are three different ways it could be broken, which we will elaborate on further in the main text, each leading to different results for anomaly mediation. First, no-scale breaking can refer to a stabilization mechanism for the radion. This generates a potential that sets the size of the extra dimension, $r$ (the scalar component of $\Sigma$) so that the radion is not a flat direction. This does not change the equation of motion for $F_\Sigma$, however, so anomaly mediation is suppressed and the potential is minimized at $V=0$. No-scale breaking can also refer to generating a non-trivial equation of motion for $F_\Sigma$ through superpotential terms $W(\Sigma)$ that depend on $\Sigma$. This leads to a potential for the radion and opens up the possibility of a nonzero $F_C$ and an AdS$_4$ minimum, where $F_C$ will be proportional to $W' (\Sigma)$. The kinetic term for $\Sigma$ in this case still remains of the no-scale form. Finally, no-scale breaking can refer to higher-order (loop) corrections to the K\"ahler potential, which generate an $|F_\Sigma|^2$ term in the Lagrangian. This leads to a 4d AdS$_4$ scale which  is loop suppressed, being proportional to the coefficient of the $|F_\Sigma|^2$ term. We consider each of these cases in turn and determine the different consequences for SUSY-breaking, anomaly mediation and the derivation of the 4d EFT. The expressions for $F_\Sigma$ and the 5d compensator $\Fcp$ are summarized in Table~\ref{tab:F_summary} for each of the different models we consider.
\\

\noindent
Our analysis reveals several unexpected features not evident from a purely 4d analysis. These are:
\begin{itemize}
    \item In general one cannot derive the low-energy theory simply by integrating the superpotential over the fifth dimension, as has frequently been assumed. In order to obtain the correct 4d theory, one needs to first derive the potential in the 5d theory by solving for the auxiliary fields before integrating over the extra dimension to get the 4d potential. The effective superpotential, $W_{\rm eff}$, and K\"ahler potential can then be chosen to reproduce the potential and kinetic terms derived from the 5d theory. As the superpotential is merely a device to construct the potential in a supersymmetric way, this procedure agrees with the conventional procedure for dimensional reduction in non-supersymmetric theories and supersymmetric models in string theory, where the matching is done at the level of the potential.

    \item In particular, the 4d K\"ahler potential in the case of a warped extra dimension does not reflect the no-scale structure of the bulk theory. This means there is not a simple relationship between superpotential terms in the 5d vs 4d picture. For example, as is known in no-scale models, a constant superpotential in the 5d model breaks supersymmetry by turning on $F_\Sigma$, but does not contribute to the potential. In the warped 4d EFT, which does not have a no-scale K\"ahler potential, a constant superpotential preserves supersymmetry and leads to a constant negative energy density. This simple example shows that deriving the effective 4d superpotential, $W_{\rm eff}$, is more complicated than simply integrating the 5d superpotential. We demonstrate how to derive $W_{\rm eff}$ for the models we consider.

   \item Anomaly mediation is associated with the compensator in the higher-dimensional theory, which can in principle be sourced by a superpotential in the bulk or on the boundary. In the four-dimensional theory, the negative energy that cancels any positive energy from supersymmetry breaking is directly related to this same compensator, connecting anomaly mediation to the scale of supersymmetry breaking. However, we find that in 5d a negative energy density is sourced only by a superpotential in the bulk. This implies the two roles of the compensator (generating anomaly-mediated masses and the negative energy density) can in principle be decoupled in the 5d theory.

    \item This allows for two different types of anomaly-mediated masses in the 4d theory. There are masses that can be derived from within the effective theory, which come from an $F_C$ sourced by a bulk superpotential, and purely 5d contributions that must be added as SUSY-breaking matching contributions to the 4d EFT. Boundary superpotentials can generate anomaly-mediated masses for fields localized on the brane, but these should be understood as originating in five dimensions. Anomaly-mediated masses that arise in the 4d theory and take the universal form indicated by the 4d analysis are associated only with a bulk superpotential.

    \item The anomaly-mediated masses for fields on the IR brane are warped down relative to the masses on the UV brane, as has been argued in the past from a 4d perspective (see, e.g.~\cite{Luty:2000ec, Luty:2002ff,Arkani-Hamed:2004zhs,DEramo:2012vvz}). This leads to a hierarchy in the anomaly-mediated masses for fields on the  branes. We find the warping is true for $F_C$ that is sourced by a bulk superpotential, with IR masses warped down. The warping is also present for localized anomaly-mediated masses from boundary superpotentials, as was suggested in ref.~\cite{Luty:2002ff}, although we find the masses also depend on the type of no-scale breaking. 
    
    \item When there are superpotentials on the boundary, there will be additional sources of supersymmetry-breaking not readily captured by the supersymmetric effective theory of a warped geometry. In addition to the localized anomaly-mediated masses just discussed, these include  contributions from the radion $F$-term, $F_\Sigma$. $F_\Sigma$ can also originate from a bulk superpotential when no-scale is either unbroken or broken by loop corrections.
   
    \item We also find that when the no-scale structure is broken by loop corrections, boundary superpotentials  can act as sources for bulk fields. Boundary superpotential terms acting as sources for bulk fields has been discussed in the past but in most cases the source terms have been field dependent~\cite{Arkani-Hamed:1998lzu,Arkani-Hamed:1999uvg, Gherghetta:2001sa, Marti:2001iw, Arkani-Hamed:2001vvu, Maru:2003mq}. Refs.~\cite{Bagger:2001qi,Bagger:2001ep, Bagger:2002rw, Bagger:2003dy} also used a different formalism to ours to show that constant superpotentials on the boundaries could generate twisted boundary conditions for bulk fermions and lead to SUSY-breaking via the Scherk-Schwarz mechanism~\cite{Scherk:1978ta}. In this case as well, the boundary superpotentials do not trivially reduce to constant superpotentials in the 4d theory.
\end{itemize}

To summarize, viable (flat space) models in which SUSY is broken in a hidden sector require  negative energy density before including SUSY-breaking terms. 
In the 5d setup negative energy requires a breaking of the no-scale form of the K\"ahler potential in the 5d theory and a constant superpotential in the bulk. This result disagrees with the apparent 4d theory obtained by integrating the bulk superpotential over the extra dimensions, in which it appears that boundary superpotentials play the same role as a superpotential in the bulk. Nonetheless, even though they do not contribute to a negative energy density, boundary superpotentials can contribute to SUSY-breaking masses that cannot be derived from within the 4d theory, and act as sources for bulk fields.
\\

This paper is organized as follows: in section~\ref{sec:review} we discuss anomaly mediation in 4d, and show how sequestering is naturally achieved if there are extra dimensions. In section~\ref{sec:5d_issues} we present simple models that highlight some of the issues that arise when considering supersymmetry in extra dimensions. In section~\ref{sec:setup}, we discuss the formalism for supergravity (SUGRA) on 5d orbifolds. In section~\ref{sec:hypermultiplets} we discuss theories stabilized by hypermultiplets via the supersymmetric generalization of the Goldberger-Wise mechanism. We show how this can stabilize the extra dimension, but will not lead to an AdS$_4$ solution due to the no-scale structure in the bulk. In section~\ref{sec:beta_models}  we show how an AdS$_4$ solution can be generated if the no-scale structure is broken by loop corrections and there is a constant superpotential in the bulk. We also highlight the role of boundary superpotentials, which lead to SUSY-breaking masses for fields localized on the branes and can act as sources for bulk fields, but do not lead to an AdS$_4$ solution. In section~\ref{sec:condensates} we consider models stabilized by a bulk gaugino condensate, where the no-scale structure is broken because the superpotential depends on $\Sigma$. We highlight cases which lead to 4d AdS$_4$ solutions before SUSY is broken on the branes. We also discuss the connection of our 5d model to the KKLT scenario. In section~\ref{sec:AMSB} we discuss how anomaly mediation works in these models, and show how a carefully derived effective theory reproduces the results of the 5d theory. Throughout our work we encounter many subtleties when deriving 4d effective theories from the various 5d models; in section~\ref{sec:EFT_summary} we summarize and review these issues and their resolution. In section~\ref{sec:conclusions} we conclude and present our results.

\subsection{Comment on Notation}

Before we begin our main discussion, we clarify some of the notation we use throughout this work, for use as a reference. We discuss separately the notation used when discussing the 4d and 5d theories. In particular, up to this point we have used $F_C$ to refer to a generic compensator $F$-term, but will now be specific about referring to a 4d or 5d compensator. 
\\

\noindent
\textbf{4d Quantities}
\begin{itemize}
    \item The compensator chiral multiplet is defined as:
    \begin{align}
        \C_4 = \MP + \theta^2 \Fcf \, ,
    \end{align}
    Note we work in conventions where the compensator is dimensionful, so fix the scalar component of $\C_4$ to be equal to $\MP$.
    \item There are two parameterizations of the radion we use when studying extra dimensional models within a 4d EFT. Ignoring the fermionic components, for a flat extra dimension we use
    \begin{align}
        \Sigma_4 = \frac{r}{2\langle r \rangle} +\theta^2 F_{\Sigma_4} \, ,
    \end{align}
    where the subscript `$4$' distinguishes the 4d radion from the radion in the 5d theory, $\Sigma$. For a warped extra dimension we use:
    \begin{align}
        \rho = e^{-k\pi r} + \theta^2 F_\rho \, .
    \end{align}
    \item We refer to a generic superpotential in the 4d theory as $W$. When we derive an effective superpotential by matching a 4d SUGRA theory to the 5d model we use $W_{\rm eff}$. $W_{\text{na\"ive 4d}}$ refers to a na\"ively derived 4d superpotential which is given by integrating the superpotential over the extra dimension (which we will show is not a valid procedure in general).
\end{itemize}

\noindent
\textbf{5d Quantities}
\begin{itemize}
    \item In 5d there is a compensator hypermultiplet, which splits into two chiral multiplets with opposite charge under the $\mathbb{Z}_2$ orbifold parity. We label each compensator by their parity:
    \begin{align}
        \C_\pm = C_\pm + \theta^2 \Fcpm \, ,
    \end{align}
    using the boldface $\C_\pm$ to refer to the full superfield and $C_\pm$ for the scalar component. Similarly, physical bulk hypermultiplets split into two chiral multiplets:
    \begin{align}
        \Phi_\pm = \varphi_\pm + \theta \chi_{R, \, \pm} + \theta^2 \Fpm \, ,
    \end{align}
    which we also label by their orbifold parity.
    \item In our analysis, the radion, which is a field that depends only on the 4d co-ordinates, appears in both the 5d and the 4d theory. As is shown in~\cite{Lust:2025vyz} one can find the value of the radion by solving the 5d equations of motion or alternatively, by minimizing in the 4d potential which we ultimately do here. In the 5d theory the radion is defined as (ignoring fermionic components)
    \begin{align}
        \Sigma = \frac{M_5r}{2} + \theta^2 F_{\Sigma} \, .
    \end{align}

    \item A constant, field independent superpotential in the bulk is denoted by $W_{\rm bulk}$, and constant superpotentials on the boundaries at $y=0, \, \pi$ by $W_0, \, W_\pi$. $W_\lambda$ refers to the $\Sigma$-dependent superpotential generated by a gaugino condensate.
    \item In the 5d setup many of the fields and parameters are odd under the $\mathbb{Z}_2$ orbifold parity. It is sometimes useful to define the function $\Theta$
    \begin{align}
        \Theta(y) = \left \{ \begin{matrix} 1   &    \quad  0<y<\pi \\  - 1   &   \quad-\pi<y<0\end{matrix} \right . \, ,
    \end{align}
    which allows us to write $\mathbb{Z}_2$-odd parameters $g_{c, h}$ and fields $\varphi_-$ in terms of $\mathbb{Z}_2$-even quantities $\widetilde{g}_{c, h}$ and $\widetilde{\varphi}_-$:
    \begin{align}
        &g_{c, h} = \Theta(y) \, \widetilde{g}_{c, h} \, ,
        && \varphi_- =  \Theta(y) \, \widetilde{\varphi}_- \, .
    \end{align}
\end{itemize}

\section{Anomaly Mediation Review}

\label{sec:review}

In this section, we review the derivation of anomaly-mediated masses in four dimensions. 
Being a gravitational effect, it is present at some level in all realistic supersymmetric models~\cite{Randall:1998uk, Giudice:1998xp}. Anomaly-mediated mass splittings are proportional to the $F$-term of the conformal compensator, $\Fcf$, which is typically non-zero whenever supersymmetry is broken in Minkowski space. The spectrum of sparticle masses is then determined by $\Fcf$ and the $\beta$-functions of the low energy theory. Anomaly-mediated models are therefore attractive for both their generality and predictivity.\footnote{The insensitivity of anomaly mediation to UV physics has also allowed for recent progress in connecting results from supersymmetric QCD~\cite{Seiberg:1994bz, Seiberg:1994pq} to QCD-like theories with broken supersymmetry~\cite{Murayama:2021xfj, Csaki:2021xhi, Csaki:2021aqv, Csaki:2021jax, Kondo:2025njf, Csaki:2025fkh, Gherghetta:2025kff}.}

Our starting point is the supergravity Lagrangian describing chiral superfields $Q_i$ coupled to vector superfields $V$:
\begin{align}
    \frac{\mathcal{L}}{\sqrt{-g}} &= \dfour |\C_4|^2 f \left( Q_i^\dag, e^{-V} Q_i\right) + \left[ \dsq \left( \C_4^3 W (Q_i) + \tau (Q_i) \mathcal{W}_\alpha \mathcal{W}^\alpha \right) + h.c. \right] \, ,
    \label{eq:4daction}
\end{align}
where $\tau$ is the gauge coupling, $\mathcal{W}_\alpha$ the gauge field strength, $W$ is the superpotential and $f$ is related to the K\"ahler potential $K$ by
\begin{align}
    f = -3 e^{-K/3} \, .
\end{align}
Notice that as in ref.~\cite{Randall:1998uk}, we introduce the function $f$ since it more clearly reflects when a theory is sequestered. Throughout this work we use conventions where $\C_4$ is dimensionful, and $f$, $K$ and $W$ are all dimensionless. To go to the parameterization more typically used in 4d models where $\C_4$ is dimensionless, each of $\C_4$, $f$, $K$ and $W$ quantity should be rescaled by appropriate powers of $\MP$. This convention will be helpful when we move to the 5d theory, where the $M_5$ dependence is implicit in the compensator dependence.

The conformal compensator, $\C_4$, is a spurion superfield that is introduced to formally restore the scale invariance of the theory. Scale invariance is then explicitly broken by the gauge-fixing
\begin{align}
    \C_4 = \MP + \theta^2 \Fcf \,
    \label{eq:4dcompensator}
\end{align}
As we will show below, $\Fcf$ is the only source of negative energy density in the potential, so is typically the same order as the SUSY-breaking $F$-terms if we are to end with a theory in Minkowski space after SUSY-breaking. A non-zero $\Fcf$ does not necessarily break supersymmetry, in contrast to the $F$-components of physical chiral multiplets, and is responsible for generating the mass splittings in AdS space required for the theory to preserve supersymmetry~\cite{Gripaios:2008rg, Rattazzi:2009ux, DEramo:2012vvz, DEramo:2013dzi}. In Minkowski space, however, $\Fcf$ does contribute to supersymmetry-breaking masses.

For now we focus on deriving the effective potential for the scalar components of the $Q_i$, neglecting the fermions and gauge fields. Solving for the $F$-terms gives
\begin{align}
    F_i^\dag &= - \MP e^{K/3 } (K^{-1})_{ij} \left( \frac{\partial W}{\partial q_j} +  W 
    \frac{\partial K}{\partial q_j} \right) \, ,
    \\
    \Fcf^\dag &= \MP^2 W e^{K/3} + \frac{\MP}{3} \frac{\partial K}{\partial q_j}F^\dag_j \, ,
    \label{eq:FC_4d}
\end{align}
where $K_{ij}$ is the K\"ahler metric
\begin{align}
    K^{ij} = \frac{\partial^2 K}{\partial q^\dag_i \partial q_j} \, .
    \label{eq:Kahler_metric}
\end{align}
Plugging this into~\eqref{eq:4daction} and performing a Weyl rescaling to go the Einstein frame gives the following Lagrangian:
\begin{align}
    \frac{\mathcal{L}}{\sqrt{-g}} &= \MP^2 K^{ij} (\partial q_i^\dag)(\partial q_j) - V_F( q_i^\dag, q_i) -  V_D( q_i^\dag, q_i) \, ,
\end{align}
where $V_F$ is the potential from the $F$-terms and $V_D$ the $D$-term potential coming from the couplings to gauge fields. These are given by
\begin{align}
    V_D( q_i^\dag, q_i) &= \frac12 \sum_a D_a^2 = \MP^4 \sum_a \frac{g_a^2}{2} \left( \frac{\partial K}{\partial q_i} t^a q_i \right)^2 \, ,
    \nonumber
    \\
    V_F( q_i^\dag, q_i) 
    &= \MP^4 e^{K} \left[  (K^{-1})_{ij} \left( \frac{\partial W^\dag}{\partial q^\dag_i} + W^\dag\frac{\partial K}{\partial q^\dag_i} \right) 
    \left( \frac{\partial W}{\partial q_j} + W  \frac{\partial K}{\partial q_j} \right)  - 3 |W|^2  \right] \, ,
    \label{eq:eff_pot}
\end{align}
where the sum over $a$ in the first line is over the gauge groups in the theory. 

Here we see that $V_D$ is strictly positive and the only negative contribution to $V_F$ comes from $W$, which in turn sources $\Fcf$. Tuning the cosmological constant to zero after SUSY breaking relates $W$ to the SUSY-breaking energy density, regardless if it comes from $F$- or $D$-term breaking. This means that at the minimum:
\begin{align}
    |W|^2 = \frac{e^{-2K/3}}{3\MP^2} K^{ij} F_i^\dag F_j +  \frac{e^{-K}}{3\MP^4}\sum_a D_a^2 \, .
\end{align}
While $W$ is not responsible for breaking supersymmetry itself, it must be nonzero in any real-world model of SUSY breaking. This will in turn lead to $\Fcf \sim \MP^2 W e^{K/3} \sim \sqrt{K^{ij}F_i^\dag F_j}$ in most models of SUSY breaking -- with the exception being no-scale models, where the K\"ahler potential is such that the two terms in~\eqref{eq:FC_4d} cancel each other.

\subsection{Anomaly-mediated Masses}

\label{subsec:masses}

Anomaly mediation refers to the unavoidable gravitational contribution to the mass splittings of superpartners. It is most readily derived from a nonzero $\Fcf$, although anomaly-mediated mass terms in non-sequestered theories can also be derived without reference to a compensator~\cite{Dine:2007me}. 

In the absence of explicit mass terms, the theory is classically scale-invariant and $\C_4$ doesn't couple to other fields at tree level. This is apparent after rescaling the fields in~\eqref{eq:4daction} to remove factors of $\C_4$ to leave canonical kinetic terms. Regardless, $\C_4$ does couple at loop level, appearing along with the cutoff scale in loop corrections to formally restore scale invariance. Focusing on the gauge coupling $\tau$, this implies that the loop correction to $\tau$ is (ignoring any $\theta$-angle)
\begin{align}
    \tau = \frac{1}{g_0^2} + 2b \log \left( \frac{\mu \MP}{\Lambda_\uv \C_4 }\right) \, , 
    \label{eq:RG_tau}
\end{align}
where $\Lambda_\uv $ is the UV cutoff and $b$ is the 1-loop coefficient in the $\beta$-function for $g$:
\begin{align}
    \beta(g) = \frac{dg}{d \log(\mu)} = - b g^3 \, .
\end{align}
Substituting equation~\eqref{eq:RG_tau} back into~\eqref{eq:4daction} we find that there is now a mass term for the gauginos which takes the form\footnote{In a theory where $\Fcf \neq 0$ but SUSY is preserved, this contribution to the mass is cancelled by a counterterm on the AdS$_4$ boundary~\cite{Gripaios:2008rg, Rattazzi:2009ux}.} 
\begin{align}
    &V_{\lambda, \, \rm mass} = - \frac{\beta(g)}{2g} \frac{\Fcf}{\MP} \lambda \lambda \, .
    \label{eq:mass_gaugino}
\end{align}
This contribution to the gaugino mass is independent of any couplings between $\lambda$ and the SUSY-breaking sector, and is calculable entirely within the IR theory.

While we have focused on the gaugino mass as a specific example, the appearance of mass terms which are given by in terms of $\beta$-functions, anomalous dimensions, and $\Fcf$ is generic. Despite the many reasonable features of the spectrum, it is well known that slepton mass squareds can be negative. We will comment later on how this can be naturally resolved in a higher-dimensional context. 

\subsection{Sequestering}

\label{subsec:sequestering}

If there are direct couplings between the SUSY breaking sector and the MSSM fields, they generally dominate over the anomaly-mediated contributions. For example, if there is a Planck-suppressed coupling between a hidden sector field $X$ and the SM field $Q$,
\begin{align}
    \Delta \mathcal{L} \sim \dfour \frac{1}{\MP^2} |X|^2 |Q|^2 \, ,
    \label{eq:L_dim6}
\end{align}
this will lead to scalar masses of order 
\begin{align}
    \Delta m_q^2 \sim \frac{|F_X|^2}{\MP^2} \, ,
\end{align}
which is generically larger than the anomaly- mediated contribution, which is suppressed further by~$\beta(g)$. Anomaly mediation is typically the dominant contribution to SUSY breaking only when the K\"ahler potential and superpotential have a sequestered form 
\begin{align}
    & f = f_{\rm vis} + f_{\rm hid} \, ,
    && W = W_{\rm vis} + W_{\rm hid} \, ,
    \label{eq:sequestering}
\end{align}
i.e. any direct couplings between the MSSM and the hidden sector are suppressed by more than powers of $1/\MP$.

From a purely 4d perspective a sequestered form of the Lagrangian is difficult to justify due to terms like~\eqref{eq:L_dim6}, but can be natural in higher-dimensional theories in which the visible and hidden sectors are physically separated in the higher dimensional space. In this case (and in the absence of light fields in the bulk) the couplings between sectors is suppressed by the size of the extra dimension, justifying the use of Lagrangians of the sequestered type.

Despite this connection to extra dimensional models, most analyses of anomaly mediation are done in the 4d effective theory. In this work we aim to complete the study of sequestered anomaly mediation with a full 5d analysis. In the next section, in anticipation of the remainder of the paper, we illustrate why such a derivation can be subtle.

\section{Issues With the 5d Theory}

\label{sec:5d_issues}

In this section we illustrate some of the puzzling issues that arise when considering supersymmetry in extra dimensions. In subsection~\ref{subsec:EFT_comparison} we compare the 4d effective theories of flat and warped extra dimensions, and show that if the 4d EFT is not carefully derived, taking the AdS$_5$ scale, $k$, to zero is a singular limit in the warped EFT. We then consider a simplified 5d model in~\ref{subsec:toy} in which we highlight the need for breaking the no-scale form of the K\"ahler potential to generate a non-zero potential. We show how this breaking can lead to badly singular terms in the potential when boundary superpotentials are included. In the remainder of the paper we will show how a careful derivation of the effective theory resolves all the above issues.

\subsection{4d EFTs of Extra Dimensions (Flat vs. Warped)}

\label{subsec:EFT_comparison}

Before addressing the 5d models, we first highlight an issue already apparent at the level of the 4d EFTs when comparing flat and warped extra dimensions. In both cases there is a radion that parametrizes the radius $r$ of the extra dimension. For a flat extra dimension, the radion is parametrized by the chiral multiplet
\begin{align}
    \Sigma_4 = \frac{r}{2\langle r \rangle} +\theta^2 F_{\Sigma_4} \, ,
\end{align}
where we have dropped the fermionic component. The subscript `$4$' is chosen to distinguish it from the related field $\Sigma$ that will appear in the 5d setup. For a warped extra dimension with AdS$_5$ scale $k$, a convenient parametrization for the radion chiral multiplet is (again dropping the fermionic partner)
\begin{align}
    \rho = e^{-k\pi r} + \theta^2 F_\rho \, .
\end{align}

In both cases, the K\"ahler function $f$ can be chosen to reproduce the kinetic term for the radion and the 4d Planck scale, which is given by $\MP^2= 2\pi M_5^3\langle r \rangle, \, (M_5^3/k)$ for the flat (warped) case, respectively. Doing so leads to:
\begin{align}
    &f_{\rm flat} \left(\Sigma_4, \Sigma_4^\dag\right) = -3 \left( \Sigma_4+ \Sigma_4^\dag \right) \, ,
    && f_{\rm warped} \left( \rho, \rho^\dag \right) = -3\left( 1 - \left | \rho \right |^2 \right) \, .
    \label{eq:f_EFTs}
\end{align}
We can take the scalar component of the K\"ahler term for the warped case and expand for small $k$
\begin{align}
    -\frac13 |\C_4|^2 f_{\rm warped} &= \MP^2 \left( 1 - \left | \rho \right |^2 \right)
     = \frac{M_5^3}{k} \left( 1 - e^{-2 \pi k r} \right) 
   \underset{k \to 0}{\rightarrow}  2 \pi M_5^3 \langle r \rangle \, .
\end{align}  
So this reduces to $ -\frac13|\C_4|^2 f_{\rm flat}$ for the flat case as we take $k \to 0$, which makes sense from a 5d perspective, where $k$ is an arbitrary parameter which can be taken to zero smoothly.

However, this is no longer true when we simply add a constant, $k$-independent, superpotential in both theories. Taking $W(Q_i) = W $ to be constant in equation~\eqref{eq:4daction}, we find that the potential and $F$-terms for the flat case are: 
\begin{align}
     &F_{\Sigma_4} = \MP W\, ,
     &&\Fcf = 0\, ,
     &&&V(r) = 0\, .
\end{align}
So the superpotential leads to SUSY-breaking from $F_{\Sigma_4}$ while $\Fcf$ vanishes. The potential also vanishes, as $r$ is a flat direction. This is the well-studied `no-scale' SUSY-breaking scenario~\cite{Cremmer:1983bf, Ellis:1984kd, Lahanas:1986uc}, where a constant superpotential breaks SUSY without generating a potential. 

In the warped models the situation is very different, however. The $F$-terms and potentials in this case are:
\begin{align}
    &F_\rho = -\MP W \rho \, ,
    &&\Fcf = \MP^2 W\, ,
    &&&V(\rho) = -\frac{3\MP^4 |W|^2}{\left(1- \left | \rho \right |^2\right)^2} \, .
\end{align}
In this case the potential is a runaway with a maximum at $\rho \to 0$ (as we have not included a stabilization mechanism). The potential is also strictly negative and at the maximum $\rho=0$ we have unbroken supersymmetry in AdS$_4$. $V(\rho)$  diverges in the $k\to 0$ limit, rather than reducing to the flat space solution.

This discrepancy presents a puzzle, as we expect a smooth $k\to 0$ limit from the 5d theory where nothing special happens in this limit. This also has significant implications for anomaly mediation, as for anomaly mediation to work we want to start in AdS$_4$ with $\Fcf \neq 0$ before adding a SUSY-breaking sector to end in flat space. In the flat EFT, SUSY is already broken and the potential gives flat space with $\Fcf=0$ (although this will be modified when loop corrections are considered). In this simple model, adding a SUSY breaking sector will end in de-Sitter space and there will be no anomaly-mediated masses as $\Fcf =0$. In contrast, the potential in the warped case is always negative with $\Fcf\neq 0$, so looks like a promising starting point for anomaly-mediated models, with the caveat that at this stage the potential is unbounded as we haven't included a stabilizing sector. 

In later sections we will find that the 5d SUGRA theory has a no-scale structure similar to the 4d EFT of a flat extra dimension. The K\"ahler term $f_{\rm warped}$ does not reflect this, which makes the matching of the superpotential from the 5d to the 4d model more subtle. For example, a constant superpotential $W$ in 5d does not translate into a constant superpotential in 4d if the extra dimension is warped. The 4d EFT for warped models is constructed by first deriving the full potential in the 5d theory, before constructing a 4d effective superpotential, $W_{\rm eff}$, that reproduces this potential. We will see that when this is done carefully, the effective theory for the warped model has a smooth $k\to 0$ limit and this apparent discrepancy is resolved.

\subsection{Singular Terms in a Simple 5d Model} 

\label{subsec:toy}

In this section we consider a simple 5d model in which we highlight some of the issues that will arise in more complete models. In particular, the no-scale structure of the bulk must be broken in order to generate a potential, but doing so can lead to delta-function squared singularities in the potential when boundary superpotentials are included. For now, we  present the issues in simple examples and comment on similar problems that have previously been encountered in the literature. In the following sections we will set up a more complete theory of supersymmetry in extra dimensions and show how, when treated carefully, these singularities can be resolved by treating the boundary terms as sources for bulk fields. 

We consider two chiral superfields: a compensator $\C_+$ and a radion $\Sigma$, both of which are part of the gravitational sector in 5d. We also include constant superpotentials both on the branes and in the bulk, and ignore any warping of the extra dimension at this stage. In 5d the conformal compensator has weight $3/2$, so the powers of $\C_+$ that appear are different to those of equation~\eqref{eq:4daction}. Again we  take $\C_+$ to be dimensionful as this will be convenient when we study the full theory in section~\ref{sec:setup}. $\Sigma$ and $\C_+$ are given by
\begin{align}
    &\Sigma = \frac{rM_5}{2} + \theta^2 F_\Sigma \, ,
    &\C_+ = M_5^{3/2} + \theta^2 \Fcp \, ,
\end{align}
and the Lagrangian describing this theory is
\begin{align}
    \frac{\mathcal
    {L}_{\rm toy \ model}}{\sqrt{-g}} &= \int dy \left \{ - 3 \dfour \, (\Sigma+\Sigma^\dag) |\C_+|^{4/3} 
    - 2 \dsq \, \C_+^2 \left[ \frac{M_5 r}{2} W_{\rm bulk} + \delta(y) W_{0} + \delta(y -\pi) W_{\pi} \right] \right \} \, ,
    \label{eq:ltoy}
\end{align}
where $y \in [0, \pi]$ is the angular parameterization of the fifth coordinate. The relative factor of $M_5 r$ between the bulk and boundary superpotentials is needed  for the correct $r$ dependence of the bulk potential, as we have used the angular co-ordinate $y$.

The equation of motion for $F_\Sigma$ is:
\begin{align}
    \frac{1}{\sqrt{-g}} \frac{\delta \mathcal{L}_{\rm toy \ model}}{\delta F_\Sigma} &= - 2 M_5^{1/2} \bFcp  = 0 \, , 
\end{align} 
which sets $\Fcp$ to zero. As the potential is proportional to $\Fcp$, it vanishes identically. As in the 4d case, this result is due to the no-scale form of the K\"ahler potential~\cite{Ellis:1984kd, Lahanas:1986uc}
\begin{align}
    K = - 3 \log(\Sigma+\Sigma^\dag) \, .
    \label{eq:K_noscale}
\end{align}
Clearly the no-scale form must be broken in 5d order to generate a potential. This no-scale form of the 5d theory persists when a warp factor is included, as we consider in section~\ref{sec:setup}.

One possible source of no-scale breaking loop-corrections, which generate a K\"ahler potential term of the form~\cite{Rattazzi:2003rj, Gregoire:2004nn, vonGersdorff:2005bf}
\begin{align}
    \frac{1}{\sqrt{-g}} \, \Delta \mathcal{L}_{\rm toy \ model} &=
    - \frac{1}{3} \beta \int dy \dfour \, (\Sigma+\Sigma^\dag)^{-3} |\C_+|^{4/3} \, ,
    \label{eq:delta_ltoy}
\end{align}
where $\beta$ parametrizes the size of these corrections.\footnote{The precise form of this term is not important, we simply want a correction that allows for a non-zero potential. However, this term is the expected correction form coming from loops in 5d~\cite{Rattazzi:2003rj, Gregoire:2004nn, vonGersdorff:2005bf}; similar corrections are expected in string theory but scale as $\sim (\Sigma+\Sigma^\dag)^{-2}$~\cite{vonGersdorff:2005bf, Berg:2005ja, Berg:2005yu, Gao:2022uop}.}
After including this additional term, to leading order in $\beta$ the $F$'s are given by
\begin{align}
    \Fcp^\dag &= - \frac{2 \beta}{M_5^{5/2} r^5} \left( M_5 r \, W_{\rm bulk} + 2 W_{0} \delta(y) + 2 W_{\pi} \delta(y -\pi) \right) \, ,  \nonumber \\ 
    F_\Sigma^\dag &= - M_5 \left( M_5 r \, W_{\rm bulk} + 2 W_{0} \delta(y) + 2 W_{\pi} \delta(y -\pi) \right)  +\mathcal{O}(\beta)  \, .
    \label{eq:F_eqns_toy}
\end{align}
While the potential no longer vanishes, it contains badly singular terms proportional to the squares of $\delta$-functions. These terms are
\begin{align}
     \frac{\mathcal{L}_{\rm singular}}{\sqrt{-g}} = 
     \frac{16 \beta }{M_5^{5/2} r^5} \left( \left | W_0 \delta(y)\right |^2 + \left | W_\pi \delta(y -\pi)\right |^2 \right) \, .
     \label{eq:L_sing}
\end{align}
In the absence of $\beta$, these problems wouldn't arise, but in this case there would also be no anomaly mediation as $\Fcp$ would be zero. This problem, though it can be made less severe, is not solved by including a finite width for the branes. In this case the delta squared terms would be replaced by terms which scale like the inverse width of the brane, so these terms are not readily regulated away.

One might be tempted to integrate over $y$ first (at the level of the superpotential) to remove the $\delta$-functions, in which case the $F$-terms become
\begin{align}
    \bFcp &= - \frac{2 \beta}{\pi M_5^{5/2} r^5} \left( \pi M_5 r \, W_{\rm bulk} + W_{0} + W_{\pi}  \right) \, ,  \nonumber \\ 
    F_\Sigma^\dag &= - \frac{M_5}{\pi} \left( \pi M_5 r \, W_{\rm bulk} + W_{0} + W_{\pi} \right)  +\mathcal{O}(\beta)  \, ,
\end{align}
and the Lagrangian trivially contains no singular terms. This is also implicit in some papers which work in the 4d effective theory and distinguish between superpotentials coming from the UV and IR branes. 

However, this approach of integrating the superpotential over $y$ creates some conceptual inconsistencies. In non-supersymmetric models, we first find the zero-mode solutions for bulk fields before integrating the Lagrangian over the extra dimension to derive the 4d theory. The supersymmetric case should follow the same logic, so the equations of motion, including those for the auxiliary fields, should be solved at the 5d level and the full Lagrangian (rather than the superpotential) be integrated over $y$.

Another approach would be to add counterterms to cancel the singularities. Even if such delta squared terms were present in a supersymmetric Lagrangian, being nonholomorphic, they would have to appear in the K\"ahler potential. They would therefore scale differently to the superpotential terms and be independently renormalized, so could  cancel the singular terms only at one energy scale. Singular terms in the K\"ahler potential would also lead to $\Fcp$ and $F_\Sigma$ being proportional to $ (1+\delta(y))^{-1}$, leading to an infinite series of singular terms in the potential. 

In a related scenario, ref.~\cite{Mirabelli:1997aj} found $\delta(y)^2$ terms in a 5d Yang-Mills theory where the charged matter was on the boundaries. In their case, the singular term they found was required to regulate a divergence in the scattering of boundary matter via the exchange of the bulk gauge fields. This is different to our case where the singular terms arise even if the boundary matter is decoupled completely from the bulk, or not present at all. These field-independent singular terms are relevant when solving for auxiliary fields, and must cancel among themselves.

When brane-localized terms for bulk fields are added to the Lagrangian they are naturally interpreted as boundary terms for bulk fields. Derivatives of bulk fields with respect to $y$ can have singular pieces that cancel the singular terms in the potential, leading to a finite on-shell action. The difference in the supersymmetric case is that the auxiliary fields $\Fcp, \, F_\Sigma$ are non-dynamical, so can't shift to cancel the singular terms to give a finite potential. Using field-dependent boundary superpotentials in this way to source bulk hypermultiplets is well understood~\cite{Arkani-Hamed:1998lzu,Arkani-Hamed:1999uvg, Arkani-Hamed:2001vvu, Maru:2003mq}. Interestingly, we will see that constant boundary superpotentials can play a similar role to the explicit source terms, even with no explicit dependence on bulk fields. 

A consequence of this is that $W_0, \, W_\pi$ should be thought of as field-dependent terms in the effective superpotential when $\beta \neq 0$, meaning they generate positive contributions to the potential. To construct a realistic model with an AdS minimum before SUSY-breaking, the source of this negative energy density must be a superpotential in the bulk rather than on the branes. All of this will be made more explicit in sections~\ref{sec:hypermultiplets}, \ref{sec:beta_models} and~\ref{sec:condensates}, but in order to reach these conclusions we first review the details of the 5d SUGRA theory.

\section{5d Supergravity on Orbifolds}

\label{sec:setup}

In this section we review the construction of supergravity in 5d. The formalism for supergravity (SUGRA) in 5 dimensions was established by looking at the SUSY transformations of fields in the 5d $\mathcal{N}=2$ theory to determine invariant actions~\cite{Mirabelli:1997aj, Zucker:1999ej, Zucker:1999fn, Altendorfer:2000rr, Kugo:2000hn, Bergshoeff:2000zn, Falkowski:2000er, Kugo:2000af, Kugo:2002js}. This was later simplified via the construction of a superspace formalism which reproduces the results of these papers~\cite{PaccettiCorreia:2004suu, Abe:2004ar}, an approach we adopt here. These papers also discuss the extension of our model to include additional hypermultiplets or vector multiplets. This formalism has been applied in some phenomenological models~\cite{PaccettiCorreia:2004iiw, PaccettiCorreia:2005pap}, but a detailed study of higher-dimensional anomaly mediation has thus far been absent from the literature.

In order to break the $\mathcal{N}=2$ supersymmetry of the 5d theory down to $\mathcal{N}=1$, we take the extra dimension to be (topologically) an $S_1/\mathbb{Z}_2$ orbifold.
The bulk $\mathcal{N}=2$ multiplets then split into pairs of $\mathcal{N}=1$ multiplets with opposite orbifold parities, while Lagrangian terms localized to the boundaries of the extra dimension only satisfy $\mathcal{N}=1$ supersymmetry. 
The metric we consider is the warped product
\begin{align}
    ds_5^2 = G_{MN} dx^M dx^N= e^{2\sigma(y)} g_{\mu \nu} dx^\mu dx^\nu - r^2 dy^2 \, ,
\end{align}
where we use $M,N$ to refer to 5d spacetime indices, $\mu, \nu$ to refer to the 4d indices and $y$ is an angular co-ordinate which takes values in the interval $ [-\pi, \pi]$ of the 5th dimension. The orbifold symmetry acts as $y \to -y$, and we put branes at each of the fixed points $y = 0, \, \pi$, each of which can host an independent $\mathcal{N}=1$ theory. The bulk fields also come in representations of $SU(2)_U$, which is related to the $R$-symmetry of the bulk theory. We label $SU(2)_U$ indices with lower case Latin indices $i, j$. This symmetry is broken down to the $U(1)_R$ symmetry of the 4d theory, where we take the generator of this symmetry to be the $\sigma_3$ direction of $SU(2)_U$. 
We take the orbifold action on a spinor $\lambda^i$ to be~\cite{Bergshoeff:2000zn}:
\begin{align}
    \lambda^i (y) \to \Pi(\lambda) \gamma_5 (\sigma_3 )^i_j \lambda^i (-y) \, ,
    \label{eq:orbifoldaction}
\end{align}
where $\Pi(\lambda) = \pm 1 $ is the intrinsic parity of $\lambda$, and we will use subscripts $\pm$ to refer to the orbifold parities of hypermultiplets. 

\subsection{Dimensional Reduction of 5d Multiplets}

Our model consists of the gravitational multiplets and a single hypermultiplet in the bulk, with the hypermultiplet used to stabilize the extra dimension via the supersymmetric generalization~\cite{Maru:2003mq} of the Goldberger-Wise mechanism~\cite{Goldberger:1999uk, Goldberger:1999un}. 
The fields which will be most relevant to this analysis will be the compensator, a stabilizing hypermultiplet, the radion, and the $D$ term from the graviphoton. These will become clear momentarily.
\\

\noindent
The full list of SUSY multiplets in the bulk theory are:
\begin{itemize}

    \item The compensator comes in a hypermultiplet, which splits into two chiral multiplets with opposite parities 
    \begin{align}
        \C_\pm = C_\pm + \theta^2 \Fcpm \, ,
    \end{align}
    where we neglect the fermionic components and will fix the scalar components $C_\pm$ by other gauge-fixing conditions discussed in section~\ref{subsec:gauge}. We take $\C_\pm$ to have mass dimension $3/2$, noting that this is different to the 4d models discussed in section~\ref{sec:review} where the compensator is dimension one. Fixing the values the scalar components $C_\pm$ is discussed in section~\ref{subsec:gauge}.

    \item We also include a physical hypermultiplet, which splits into two 4d chiral multiplets
    \begin{align}
        \Phi_\pm = \varphi_\pm + \theta \chi_{R, \, \pm} + \theta^2 \Fpm \, ,
    \end{align}
    where the fields $\varphi_\pm$ will be used to stabilize the extra dimension.

	\item The Weyl multiplet containing the gravitational fields: 
	\begin{align}
		\left( e^N_M, \psi^i_M , B^{ij}_M, b_M, v_{MN}, \chi^i, X \right) \, .
        \label{eq:weyl-multiplet}
	\end{align}
    The only dynamical fields above are the veilbein, $ e^N_M$, and the gravitino, $\psi^i_M$~\cite{Gunaydin:1984ak}. The others are either fixed by gauge conditions ($b_M$) or are auxiliary fields to be solved for ($B^{ij}_M, v_{MN}, \chi^i, X$). $\chi^i$ and $X$ ultimately appear in the action as Lagrange multipliers which enforce gauge-fixing conditions. In our analysis we drop these fields and enforce the gauge-fixing conditions by hand, following the approach of ref.~\cite{Abe:2004ar}.

    We find it useful to make the basis change
    \begin{align}
        &B^{ik}_M \epsilon_{kj} = i \sum_{r=1}^3 B^r_M (\sigma_r)^i_j \, ,
        \label{eq:SU2basis_V}
    \end{align}
    where the $\sigma$'s are Pauli matrices and the fields $B_M^{(3)}$ and $B_M^{(1, 2)}$ have opposite parities as a result of the orbifold action~\eqref{eq:orbifoldaction}. The fields with positive parity are:
    \begin{align}
		\left( e^\nu_\mu, e^{y}_y,  \psi^1_{\mu, R},  \psi^2_{y, R} , B^{3}_\mu, B^{1, 2}_y, b_\mu, v_{\mu y}, \chi^1_R, X \right) \, ,
	\end{align}
    where $\psi$ and $\chi$ are fermionic fields and the rest are bosons. The negative parity fields are:
    \begin{align}
		\left( e^y_\mu, \psi^2_{\mu, R},  \psi^1_{y, R}, B^{3}_y, B^{1, 2}_\mu, b_y, v_{\mu \nu}, \chi^2_R \right) \, .
	\end{align}
    
    \item There is also the graviphoton multiplet:
    \begin{align}
        \left( M, W_M, \Omega^i, Y^{ij} \right) \, ,
    \end{align}
    where it is convenient to make the basis rotation~\eqref{eq:SU2basis_V} for $Y^{ij}$,
    \begin{align}
        &Y^{ik} \epsilon_{kj} = i \sum_{r=1}^3 Y^r (\sigma_r)^i_j \, .
        \label{eq:SU2basis_Y}
    \end{align}
    The gauge field $W_\mu$ has odd orbifold parity and couples to the $U(1)_R$ subgroup of $SU(2)_U$ which is generated by $\sigma_3$. This symmetry is explicitly broken by the orbifolding. 
\end{itemize}
The $y$ component of the auxiliary field $B$ from the Weyl multiplet mixes with the fields from the graviphoton multiplet to give an odd vector multiplet, $V$, which is the graviphoton multiplet, and an even chiral multiplet, $\Sigma$, which can be identified as the radion multiplet:
\begin{align}
    V &\equiv \theta \overline{\sigma}^\mu \overline{\theta} \, W_\mu + i \theta^2 \overline{\theta} \, \lambda^\dag - i \overline{\theta}^2 \theta \, \lambda + \frac{1}{2} \theta^2 \overline{\theta}^2 D \, , \nonumber \\
    \Sigma &\equiv \varphi_\Sigma + \theta \chi_\Sigma + \theta^2 F_\Sigma \, .
\end{align}
The components of $V$ and $\Sigma$ are
\begin{align}
    \lambda &\equiv 2 e^{\frac{3}{2} \sigma} \Omega^1_{R}\, , \nonumber \\
    D &\equiv -e^{2\sigma} \{ r^{-1} \partial_y M - 2Y^3 + r^{-1} \dot{\sigma} M \}\, , \nonumber \\
    \varphi_\Sigma &\equiv \frac{1}{2} \left( r M  - i W_y \right)\, , \\
    \chi_\Sigma &\equiv 2 e^{\frac{\sigma}{2}} \left( r \Omega^2_{R} - i\psi^2_{y, R} M_5^{-1/2} \right)\, , \nonumber \\
    F_\Sigma &\equiv - e^\sigma \{ (B^1_y + iB^2_y) M - i r (Y^1 + iY^2) \} \, . \nonumber 
    \label{eq:vec}
\end{align}
The above fields appear in the action in the gauge-invariant combinations 
\begin{align}
    \mathcal{V}_\Sigma &= \Sigma + \Sigma^\dag  -\partial_y V \, ,  \nonumber \\
    \mathcal{W}_\alpha &= - \frac{1}{4} \overline{D}^2 D_\alpha V = i \left( \lambda_\alpha -\theta^\beta \left( (\sigma^{\mu \nu})_{\beta\alpha} F_{\mu \nu} +i\epsilon_{\beta\alpha} D \right) + i\theta^2 \sigma^\mu \partial_\mu  \lambda^\dag_\alpha \right) \, ,
\end{align}
where $F_{\mu \nu} $ is the field strength of $W_\mu$.

To generate a warped 5d spacetime, we choose the compensators $C_\pm$ to have couplings $\pm g_c$ under the $U(1)$ gauged by the graviphoton. We will also take the stabilizing multiplets $\Phi_\pm$ to have couplings to the graviphoton $\pm g_h$, where both couplings are $\mathbb{Z}_2$-odd parameters. The couplings can be written in terms of even parameters $\widetilde{g}_c, \, \widetilde{g}_h$ by explicitly including the function $\Theta$:
\begin{align}
    &g_{c, h} = \Theta(y) \, \widetilde{g}_{c, h}
    && \Theta(y) = \left \{ \begin{matrix} 1   &    \quad  0<y<\pi \\  - 1   &   \quad-\pi<y<0\end{matrix} \right .
    \label{eq:odd_couplings}
\end{align}
Ultimately, $g_c$ will determine the bulk AdS scale, while the masses of the stabilizing fields are functions of both $g_c$ and $g_h$.

\subsection{The Bosonic 5d Action}

The action of the theory described in the previous section can be written in the superspace formalism as~\cite{Abe:2004ar}
\begin{align}
    \mathcal{L} &= \mathcal{L}_{\rm vec} + \mathcal{L}_{\rm K} + \mathcal{L}_{\rm W} \, , \nonumber \\
    \mathcal{L}_{\rm vec} &= - \left[ \dsq \frac{3}{2} \left\{ \Sigma \,  \mathcal{W}^\alpha \mathcal{W}_\alpha - \frac{1}{12} \overline D^2 \left( V D^\alpha \partial_y V - D^\alpha V \partial_y V \right) W_\alpha \right\} + \text{h.c.} \right] 
    \,, \nonumber \\
   \mathcal{L}_{\rm K} &=- e^{2\sigma} \dfour \left \{ \mathbb{W}_y^{-2} \mathcal{V}_\Sigma^3
    +2 \mathbb{W}_y \left[ \C^\dag_a \left( e^{-2g_c \sigma_3 V} \right)^a_b \C^b - \Phi^\dag_a \left( e^{-2 g_h \sigma_3 V } \right)^a_b  \Phi^b  \right] \right\}
     \,, \nonumber \\
   \mathcal{L}_{\rm W} &= e^{3\sigma} \left[ \dsq \ \Phi_a \left( \partial_y + 2 \Sigma g_h \sigma_3 \right)_b^a \Phi^b - \C_a \left( \partial_y + 2 \Sigma g_c \sigma_3 \right)_b^a \C^b + \text{h.c.} \right]\, .
    \label{eq:action}
\end{align}
For the hypermultiplets in eq.~\eqref{eq:action}, the indices $a, b$ take the values $+, -$ and are raised and lowered by the 2-d Levi-Civita tensor. Note that we are writing the Lagrangian in a manifestly $\mathcal{N}=1$ symmetric way, but the bulk supersymmetry is $\mathcal{N}=2$ (up to radiative corrections).

To simplify $\mathcal{L}_{\rm K}$ we follow the approach of ref.~\cite{PaccettiCorreia:2006rpo} and integrate out the field $\mathbb{W}_y$ to get\footnote{The notation $\mathbb{W}_y$ matches that of ref.~\cite{PaccettiCorreia:2004iiw}. In ref.~\cite{Abe:2004ar} this field is referred to as $V_T$, with the radion multiplet denoted $\Phi_S$.}
\begin{align}
   \mathcal{L}_{\rm K} &= - 3 e^{2\sigma} \dfour \ \mathcal{V}_\Sigma  \left[\C^\dag_a \left( e^{-2g_c \sigma_3 V} \right)^a_b \C^b - \Phi^\dag_a \left( e^{-2 g_h \sigma_3 V } \right)^a_b  \Phi^b \right]^{2/3}  \, .
\end{align}
In this form the no-scale structure of $\mathcal{L}_{\rm K}$ becomes evident, with the K\"ahler potential given by
\begin{align}
    K = -3 \log \mathcal{V}_\Sigma = -3 
    \log \left( \Sigma + \Sigma^\dag -\partial_y V \right) \, .
\end{align}
The equation of motion for $F_\Sigma$ then sets
\begin{align}
    C_- \Fcm + C_+ \Fcp = \varphi_+ \Fp + \varphi_- \Fm -e^\sigma g_h M_5 \varphi_+ \varphi_- \, .
    \label{eq:noscale_EOM}
\end{align}
This limits the possibility of generating a negative energy density through $\Fcpm$. The last term actually breaks supersymmetry (through $F_\Sigma$), as we will see in~\eqref{eq:Fterms}, and the compensator $F$-components must be smaller than $\Fpm$ as we expect $C_+ \gg \varphi_\pm$ -- that is field values should be less than $M_5$. Breaking the no-scale form to the K\"ahler potential will be crucial to finding an anomaly-mediated spectrum and a flat 4d theory after SUSY-breaking.

\subsubsection{Gauge Fixing}

\label{subsec:gauge}

Many of the degrees of freedom we have so far introduced are non-dynamical and must be fixed in order to explicitly break some of the extraneous symmetries of the theory. We first break the $SU(2)_U$ symmetry by setting 
\begin{align}
    C_- = 0 \, ,
    \label{eq:Cminus}
\end{align}
We could allow for $C_-$ to be nonzero, but this will lead to a non-vanishing Wilson line in the bulk. This breaks supersymmetry via the Scherk-Schwarz mechanism~\cite{Scherk:1978ta} by turning on $F_\Sigma$ proportional to the Wilson line~\cite{Kaplan:2001cg}; we do not consider this possibility in this work. In ref.~\cite{Son:2008mk}, the authors considered $C_-$ turning on in response to a SUSY-breaking perturbation to the boundary. However, $C_-$ is not a dynamical field and should be fixed to break some of the spurious symmetries of the theory -- allowing $C_-$ to respond to a perturbation effectively means working in a different theory before and after the perturbation -- in this case indicated by the non-vanishing Wilson line when $C_- \neq 0$. In any case it is important to recognize that these are auxiliary fields whose values are fixed once in the beginning of the analysis.

We can then break the (spurious) 4d dilatation symmetry  by fixing $C_+$. In order to get the correct coefficient of the Einstein-Hilbert action in the 5d theory (so we work in the Einstein frame in 5d), we  set $M=M_5$ and fix $C_+$ to be
\begin{align}
    &|C_+|^2 - |\varphi_+|^2 -|\varphi_-|^2 = M_5^3 \, ,
    \nonumber
    \\
    \implies 
    &C_+ = \sqrt{M_5^3 + |\varphi_+|^2 +|\varphi_-|^2 } \, .
    \label{eq:gaugefixing}
\end{align}
We note that the terms which go like $\partial_y M$ in the Lagrangian should be treated by integrating by parts to take the derivatives off $M$ first before fixing $M=M_5$. For the purposes of this work we also take the axion to vanish, $A_5 =0$.

Additional gauge-fixing constraints set some of the fermion fields and components of the Weyl multiplet to zero, see appendix~\ref{app:fermiongauge} for details. As we keep only the bosonic components of the chiral multiplets, the relevant conditions for our discussion are equations~\eqref{eq:Cminus} and~\eqref{eq:gaugefixing}. We also neglect the graviphoton $W_\mu$, keeping only the auxiliary field $D$ from the vector multiplet. We also assume a Lorentz-invariant background and ignore gradients in the 4d directions. 

\subsection{Supersymmetric Bulk Solutions}

We now solve the equations of motion for the auxiliary fields, and derive the conditions needed for the bulk profiles to preserve supersymmetry. The SUSY-breaking auxiliary fields are the $D$-component of the vector multiplet and the $F$-components of $\Sigma$ and $\Phi_\pm$.

\subsubsection{Gravitational Terms and the Warp Factor}

The terms in $\mathcal{L}_{\rm vec}$ and $\mathcal{L}_{\rm K}$ involving the auxiliary field, $D$, of the graviphoton are responsible for giving the Einstein-Hilbert action in the 5d theory. Hence we label these contributions to the Lagrangian as $\mathcal{L}_{\rm D}$, which is given by
\begin{align}
     \mathcal{L}_{\rm D} = -\frac{3}{2} D^2 M_5 r - 2 r e^{2 \sigma} D \left( (g_c+g_h) |\varphi_-|^2 + (g_c-g_h) |\varphi_+|^2 +M_5^2 \left(g_c M_5 + \frac{3\dot \sigma}{2r} \right)\right)\, .
\end{align}
The supersymmetric solution for the warp factor can then be derived by setting $D$ to vanish: 
\begin{align}
    D &= -\frac{M_5 e^{2\sigma}}{r} \left[  \dot \sigma + \frac{2r}{3M_5^2} \left( g_c M_5^3 + (g_c+g_h)|\varphi_-|^2 + (g_c - g_h) |\varphi_+|^2  \right) \right] = 0 \, ,
    \label{eq:D_sol}       \\
    \implies \dot \sigma &=  - \frac{2r}{3M_5^2} \left( g_c M_5^3 + (g_c+g_h)|\varphi_-|^2 + (g_c - g_h) |\varphi_+|^2  \right) \, .
    \label{eq:warp_SUSY}
\end{align}
We will see in the next section that when the fields $\varphi_\pm$ have supersymmetric background profiles, this also solves Einstein's equations.  In the limit where the backreaction from the stabilizing fields is small, $|\varphi_\pm| \ll M_5^{3/2}$, this is just the usual warp factor of the RS model, $\sigma = - kr y$, with the AdS$_5$ scale given by
\begin{align}
    k = \frac{2g_c M_5}{3} \, .
\end{align}

\subsubsection{Hypermultiplet Lagrangian}
\label{subsubsec:hypL}

The remaining Lagrangian terms can be written in terms of the $F$-components. We denote these terms $\mathcal{L}_{\rm F}$, so that the full Lagrangian is $\mathcal{L} = \mathcal{L}_{\rm F} + \mathcal{L}_{\rm D}$. These contributions are given by
\begin{align}
    \mathcal{L}_{\rm F} =& - 2r e^{2 \sigma} \left( |\Fcm|^2+|\Fcp|^2-|\Fm|^2-|\Fp|^2 - \frac{1}{3M_5^3r} \left |\bFm \varphi_- + \bFp  \varphi_+ - \bFcp C_+\right |^2  \right)  \nonumber
    \\
    & \hspace{20mm} - \frac{2e^{2 \sigma}}{M_5}\left[ F_\Sigma (\bFcp C_+ - \bFm \varphi_- - \bFp  \varphi_+ ) + h.c. \right]  \, ,  \nonumber
    \\
    & + 2 e^{3 \sigma}\bigg[ \Fcm \left(\partial_y + \frac{3\dot \sigma}{2} + g_c M_5 r \right) C_+  -2 F_\Sigma \, g_h \varphi_- \varphi_+  \nonumber
    \\
    & \hspace{15mm} + \Fp \left( \partial_y + \frac{3\dot \sigma}{2} - g_h M_5 r \right)\varphi_-- \Fm \left( \partial_y + \frac{3\dot \sigma}{2} + g_h M_5 r \right)\varphi_+  + h.c. \bigg]  \, .
\end{align}
This leads to the following solutions for the $F$-components:
\begin{align}
    \bFcp &= \frac{e^{\sigma}  \left(\varphi_- \dot \varphi_+ - \varphi_+ \dot \varphi_- \right) C_+}{M_5^3 r} \, , \nonumber
    \\
    \bFcm &= \frac{e^\sigma}{r} \left(\partial_y + \frac{3\dot \sigma}{2} + g_c M_5 r \right) C_+\, ,\nonumber
     \\
    \bFp &= \frac{e^{\sigma}}{r} \left[ - \left( \partial_y + \frac{3\dot \sigma}{2} - g_h M_5 r \right)\varphi_- 
    + \frac{\varphi_+^\dag \left(\varphi_- \dot \varphi_+ - \varphi_+ \dot \varphi_- \right)}{M_5^3} \right] \, , 
    \label{eq:Fterms}         \\
    \bFm &= \frac{e^{\sigma}}{r} \left[ \left( \partial_y + \frac{3\dot \sigma}{2} + g_h M_5 r \right)\varphi_+
    +  \frac{ \varphi_-^\dag \left(\varphi_- \dot \varphi_+ - \varphi_+ \dot \varphi_- \right)}{M_5^3} \right] \, , \nonumber
    \\
    F_\Sigma^\dag &= \frac{e^{\sigma}\left( 3\varphi_+ \dot \varphi_- - 3 \varphi_- \dot \varphi_+
    - 2 g_h M_5 r  \varphi_-  \varphi_+\right)}{3 M_5^2} \, .
    \nonumber
\end{align}
Here we can see some requirements that a supersymmetric solution must satisfy. One can check that a supersymmetric solution has only one of the fields $\varphi_\pm$ non-zero. The SUSY-preserving profiles for $\varphi_\pm $ satisfy the first order equations of motion
\begin{align}
    \left( \partial_y + \frac{3\dot \sigma}{2} \pm g_h M_5 r \right)\varphi_\pm = 0 \, .
    \label{eq:F_EOM}
\end{align}
The solutions to~\eqref{eq:F_EOM} are not the full set of solutions to the equation of motion, which is a second order differential equation. In the limit of small backreaction, the general solutions for $\varphi_\pm$ are given by
\begin{align}
    \varphi_\pm (y) = A_\pm e^{(g_c \mp g_h) y/2} + B_\pm e^{(5g_c \pm 3g_h) y/6} \, .
    \label{eq:hyper_profiles}
\end{align}
The supersymmetric profile for each of the hypermultiplets therefore corresponds to $B_+ = B_- = 0$.  The full supersymmetric solution further requires that only one of $A_\pm$ be nonzero.  

We can now check that the other auxiliary fields do not switch on in the presence of a supersymmetric background profile for either of $\varphi_\pm$. Assuming flat 4d space, the Einstein equation determining the warp factor is~\cite{DeWolfe:1999cp}
\begin{align}
    \dot \sigma^2 &= (kr)^2 + \frac{1}{6 M_5^3 k r} \left[|\dot \varphi_-|^2 + |\dot \varphi_+|^2 + r^2\left( m_-^2 |\varphi_-|^2 +  m_+^2  |\varphi_+|^2\right) \right] \, ,\label{eq:warp_einstein} 
\end{align}
where $m_\pm^2 = g_h^2 - \frac{5g_c^2}{3} \pm \frac{2g_c g_h}{3}$. Comparing \eqref{eq:warp_SUSY} to \eqref{eq:warp_einstein} it may seem surprising that we can set $D=0$ when turning on background profiles for $\varphi_\pm$. However, when $\varphi_\pm$ preserve supersymmetry (i.e. $B_+ = B_- = 0$), it can be shown that the solution to the full Einstein equation~\eqref{eq:warp_einstein} also solves the $D=0$ equation. Furthermore, $\Fcm = 0$ when we include the dependence of $C_+$ on the hypermultiplets through equation~\eqref{eq:gaugefixing} and use the supersymmetric solutions for $\varphi_\pm$. 
\\

This formalism also automatically gives the right tensions on both of the branes to give 4d flat space~\cite{PaccettiCorreia:2004suu}. We show schematically how this works here, leaving the details for appendix~\ref{app:IBP}. After substituting the solutions for the auxiliary fields we have the non-canonical kinetic terms
\begin{align}
    \mathcal{L}_{\rm F}  =  \frac{2 e^{4\sigma}}{r} \left \{ 
    \left | \left( \partial_y + \frac{3\dot \sigma}{2} + g_c M_5 r \right) C_+\right |^2 
    - \left | \left( \partial_y + \frac{3\dot \sigma}{2} - g_h M_5 r \right) \varphi_- \right |^2 
    -  \left | \left( \partial_y + \frac{3\dot \sigma}{2} + g_h M_5 r \right) \varphi_+ \right |^2 \right \} + \ldots \, ,
    \label{eq:Lkin}
\end{align}
with no explicit boundary terms. Expanding equation~\eqref{eq:Lkin} we will find terms like
\begin{align}
     - \frac{2 e^{4\sigma}}{r} (\partial_y \varphi^\dag_- ) g_c M_5 r\varphi_- + h.c. \, ,
\end{align}
which can be integrated by parts, leaving only mass terms and terms proportional to $|\partial_y \varphi_-|^2$. We remind the reader that we have taken $g_{c}$ and $g_h$ to be odd parameters, so derivatives acting on these couplings give boundary terms, ${\dot g_{c/h} = 2\left[\delta(y) - \delta(y-\pi) \right] \widetilde g_{c/h}}$. In the basis where the hypermultiplets have canonical kinetic terms the boundary terms we find are
\begin{align}
    \mathcal{L}_{\rm F} \big |_{\rm bdy} &= \left[\delta(y) - \delta(y-\pi)\right] 4 M_5e^{4\sigma} \left( \widetilde g_c|C_+|^2 - \widetilde g_h |\varphi_+|^2 - \widetilde g_h |\varphi_-|^2 \right) \, ,   \nonumber
    \\
    &= \left[\delta(y) - \delta(y-\pi)\right] 4 M_5 e^{4\sigma} \left( \widetilde g_c M_5^3 + (\widetilde g_c + \widetilde g_h) |\varphi_-|^2 + (\widetilde g_c - \widetilde g_h) |\varphi_+|^2 \right) \, ,
    \label{eq:tensions}
\end{align}
which is precisely the correct value to match the discontinuity in the warp factor (given in equation~\eqref{eq:warp_SUSY}) at the boundary.
Making the identification $2\widetilde g_c M_5 = 3k$, the tension of the branes are $T_{0/\pi} = \pm 12 M_5^3 k$, which can be recognized as the tuned brane tensions in the RS model~\cite{Randall:1999ee}.  Notice this gives both the correct tensions and also introduces new mass terms on the boundary that are not supersymmetric on their own. In refs.~\cite{Bagger:2002rw, Bagger:2003dy} it was shown that subcritical brane tensions also preserve supersymmetry, but that situation does not arise in our setup. 

This procedure of integrating by parts to put the kinetic terms into a canonical form also reveals a coupling of the hypermultipets to $\mathcal{R}$. The coefficient of $\mathcal{R}$ from the hypermultiplets is (see appendix~\ref{app:IBP})
\begin{align}
    -\frac{3re^{4\sigma}}{8} \mathcal{R} \left( |C_+|^2 -|\varphi_-|^2 -|\varphi_+|^2 \right) =  -\frac{3rM_5^3e^{4\sigma}}{8} \mathcal{R} \, .
\end{align}
This term combines with terms from the $D$-term Lagrangian to give to give the gravitational terms for 5d AdS space:
\begin{align}
    \mathcal{L}_{\rm grav} &= - \frac{M_5^3r e^{4\sigma}}{2} \left( \mathcal{R} - 12 k^2\right) \, .
    \label{eq:L_grav}
\end{align}
The choice of gauge fixing made in equation~\eqref{eq:gaugefixing} eliminates the field dependence in front of $\mathcal{R}$, so we are working in the 5d Einstein frame. $\mathcal{R} $ contains term proportional to the 4d Ricci scalar, $\mathcal{R}^{(4)}$
\begin{align}
    \mathcal{R} = e^{-2\sigma} \mathcal{R}^{(4)} + \ldots \, .
\end{align}
From this we can determine the coefficient of $\mathcal{R}^{(4)}$ to be
\begin{align}
    \mathcal{L}_{\mathcal{R}^{(4)}} = - \frac12 \int dy M_5^3 r e^{2\sigma} \mathcal{R}^{(4)} = - \frac{\MP^2}{2} \left(1 - e^{-2k \pi  r} \right)  \mathcal{R}^{(4)}\, ,
    \label{eq:4d-EH-term}
\end{align}
where we have identified the reduced Planck scale as
\begin{align}
    \MP^2 = M_5^3 /k\, .
    \label{eq:MP_4d}
\end{align}
Note that although we are working in the 5d Einstein frame, we are not in the Einstein frame in 4d due to the radion dependence in equation~\eqref{eq:4d-EH-term}.

\subsection{Quadratic Lagrangian in 5d}

The full Lagrangian in 5d, to quadratic order in the fields, is found to be (see appendix~\ref{app:IBP} for further details)
\begin{align}
    \mathcal{L}_5 =& - \frac{M_5^3 re^{4\sigma}}{2} \left( \mathcal{R} - 12 k^2 \right) 
    - 2r e^{4\sigma} \left[ r^{-2}\left( |\dot \varphi_+|^2 +  |\dot \varphi_-|\right)^2 + m_+^2 |\varphi_+|^2 + m_-^2 |\varphi_-|^2 \right] \, , \nonumber
   \\ 
   &+ 2e^{4\sigma}M_5 \left[ (\dot g_h - \dot g_c ) |\varphi_+|^2 - (\dot g_h + \dot g_c ) |\varphi_-|^2 - M_5^3 \dot g_c   \right] \, .
\end{align}
The physical masses for the scalars are
\begin{align}
    m_\pm^2 =  \frac{M_5^2}{3} \left( 3 g_h^2 + g_c^2 \pm 2g_c g_h \right) \, ,
\end{align}
where the masses contain contributions from the $F$-term Lagrangian~\eqref{eq:Lkin}, as well as terms from $\mathcal{L}_{\rm D}$. After substituting the solution for the warp factor~\eqref{eq:warp_einstein}, many of the terms drop out and we are left with a sum of the hypermultiplet $F$-terms. When integrated over $y$, this leads to the 4d effective potential for the hypermultiplets:
\begin{align}
    V_{\rm eff} &= 2 r \int_0^\pi dy e^{2\sigma(y)} \left(\left|\Fp\right |^2 + \left|\Fm \right |^2 \right) \, , \nonumber
    \\
    &= \frac{2}{r} \int_0^\pi dy e^{4\sigma(y)} \left( \left| \dot \varphi_- - M_5r (g_c +g_h) \varphi_- \right |^2 + \left| \dot \varphi_+ - M_5r (g_c -g_h) \varphi_+ \right |^2 \right) \, .
    \label{eq:Veff_general}
\end{align}

\section{Stabilization by Hypermultiplets}

\label{sec:hypermultiplets}

In this section we include field-dependent boundary terms to stabilize the extra dimension by sourcing the fields $\varphi_\pm$, realizing a supersymmetric version of the Goldberger-Wise solution. We present the 5d solution and derive the corresponding 4d effective theory, finding that the effective potential is minimized at $V=0$. In the following section, we will show that constant boundary superpotentials can also source bulk fields, but only in the presence of no-scale breaking. 

\subsection{Source Terms}

\label{subsec:sources}

To generate a bulk profile for $\varphi_-$ we include source terms in the superpotentials on each of the branes. Before supersymmetry is broken, the  boundary terms respect $\mathcal{N}=1$ SUGRA and are composed of the even-parity bulk fields and boundary-localized fields. The conformal compensator on the branes, $\C_{\rm bdy}$,  is related to the parity-even bulk compensator by
\begin{align}
    \C_{\rm bdy} = \C_+^{2/3} \, ,
    \label{eq:bdy_comp}
\end{align}
where the power of $2/3$ is because $\C_+$ has conformal weight $3/2$ while the four-dimensional compensator has weight one. Bulk chiral multiplets with even parity appear on the boundary in the combination $\Phi_+/\C_+$. We  source $\varphi_-$ by adding the boundary superpotentials\footnote{Similar boundary potentials were also considered in refs.~\cite{Arkani-Hamed:1998lzu,Arkani-Hamed:1999uvg, Arkani-Hamed:2001vvu, Son:2008mk}.} 
\begin{align} \label{eq:boundary_W}
    &W |_{y=0} = - \frac{4 J_0\Phi_+}{\C_+} \, ,
    &&W |_{y=\pi} = \frac{ 4 J_\pi \Phi_+}{\C_+} \, ,
\end{align}
where the minus sign and factors of $4$ are included for later convenience. The terms in the boundary Lagrangian then become
\begin{align}  \label{eq:boundary}
    \mathcal{L}_{\rm bdy}
    &= 4e^{3\sigma} \left[ - \delta(y) \dsq \C_+ J_0 \Phi_+
    + \delta(y - \pi) \dsq \C_+ J_\pi \Phi_+ + h.c. \right]\, , \nonumber  
    \\
    &= 4 e^{3\sigma} \left[ -\delta(y) \left( C_+J_0 \Fp + \Fcp J_0 \varphi_+ \right)
    + \delta(y - \pi) \left( C_+J_\pi \Fp + \Fcp J_\pi \varphi_+ \right) + h.c. \right]\, .
\end{align}
We set $\varphi_+ = 0$, as we will see later that this is the only value that preserves supersymmetry. 

These terms introduce a boundary term in $\Fp$, which now reads
\begin{align}
    \bFp &= - \frac{e^{\sigma}}{r} \left( \partial_y + \frac{3\dot \sigma}{2} - g_h M_5 r \right)\varphi_- + \frac{2e^{\sigma} M_5^{3/2} }{r} \left[ \delta(y) J_0 - \delta(y - \pi) J_\pi \right]  + \mathcal{O} \left(\frac{|\varphi_\pm|^3}{M_5^3} \right)\, ,
\end{align}
where we have kept only the terms linear in $\varphi_\pm$. The boundary terms can be interpreted as source terms for $\varphi_-$ arising from the $\Fp = 0$ equation of motion. To see this, we can write the odd field $\varphi_-$ as 
\begin{align}
    \varphi_- = \Theta(y) \widetilde \varphi_- \, ,
\end{align}
where $\widetilde \varphi_-$ is even. $y$ derivatives acting on $\varphi_-$ then give $\delta$-function terms:
\begin{align}
    \partial_y \varphi_- = 2 \left[ \delta(y) - \delta(y-\pi)\right] \widetilde \varphi_- + \Theta(y) \partial_y \widetilde \varphi_- \, .
\end{align}
Requiring that the coefficient of the $\delta$-functions in $\Fp$ vanishes leads to the boundary conditions for $\widetilde \varphi_-$
\begin{align}
    & \widetilde \varphi_- (0) = M_5^{3/2} J_0 \, ,
    && \widetilde \varphi_- (\pi) = M_5^{3/2} J_\pi  \, .
    \label{eq:phi-_bcs}
\end{align}

Requiring that $\bFp =0$ with the above boundary conditions then leads to the solution 
\begin{align}
    \widetilde \varphi_- (y) = M_5^{3/2} J_0 \exp \left( - \frac{3\sigma(y)}{2} + rM_5 g_h y \right) \, ,
    \label{eq:SUSY_soln}
\end{align}
where the boundary condition on the IR brane is satisfied only if
\begin{align}
    J_0 \exp \left( - \frac{3\sigma(\pi)}{2} + rM_5 g_h \pi \right) = J_\pi \, .
    \label{eq:r_crit}
\end{align}
The source terms also generate boundary terms in $\Fm, \Fcp$ and $F_\Sigma$ proportional to $\varphi_+$, which vanish after setting $\varphi_+ = 0$.

Equation~\eqref{eq:r_crit} indicates that supersymmetry is preserved only if $r$ is stabilized at a specific value given by
\begin{align}
     r = \frac{1}{(g_c + g_h)M_5\pi}\log \left( \frac{J_\pi}{J_0}\right) \, ,
     \label{eq:rSUSY}
\end{align}
where we neglect the backreaction on $\sigma$ in~\eqref{eq:rSUSY}. We will show in the next section that in the absence of additional terms, the effective potential is minimized for this value of $r$. If $r$ did deviate from this value, for example due to some energy density on the branes, then the stabilizing fields  would break supersymmetry as $\Fp$ would switch on. In the parametrization of \eqref{eq:hyper_profiles}, the coefficients of the exponential terms in the $\varphi_-$ profile are
\begin{align}
    &A_- = \frac{M_5^{3/2} \left(J_0 e^{\frac{1}{3} \pi  M_5 r (5 g_c-3 g_h)}-J_\pi\right)}{ e^{\frac{1}{3} \pi  M_5 r (5 g_c-3 g_h)}-e^{\pi  M_5 r (g_c+g_h)} } \, ,
    &&B_- = \frac{M_5^{3/2} \left(J_\pi-J_0 e^{\pi  M_5 r (g_c+g_h)}\right)}{e^{\frac{1}{3} \pi  M_5 r (5 g_c-3 g_h)}-e^{\pi  M_5 r (g_c+g_h)}} \, .
    \label{eq:C1C2-}
\end{align}
Here we see that the supersymmetric value for $r$ in~\eqref{eq:rSUSY} leads to $A_- = M_5^{3/2} J_0$ and $B_- = 0$. This is as expected since the supersymmetric $F$-term equation of motion has only one solution.

\subsection{4d Effective Theory}

\label{subsec:EFT}

To derive the effective potential we set $\varphi_+ = 0$ and substitute the solution for $\varphi_-$ defined by equations~\eqref{eq:hyper_profiles} and~\eqref{eq:C1C2-}. Expanding the action to quadratic order in $\varphi_-$ and integrating over $y$ leads to the effective potential
\begin{align}
    V_{\rm eff} = \frac{4 M_5^4 (g_c-3 g_h) \left |J_\pi-J_0 e^{\pi  M_5 r (g_c+g_h)}\right |^2}{3 \left(e^{\frac{8}{3} \pi  g_c M_5 r}-e^{2 \pi  M_5 r (g_c+g_h)}\right)} \, ,
    \label{eq:Veff_J1}
\end{align}
which vanishes at the minimum, leading to a supersymmetric solution in Minkowski space. In order to generate a large hierarchy we require $g_c + g_h \ll g_c$, in which case the potential has an overall scale $M_5^3 k e^{-4k \pi r}$ (as $2M_5 g_c  = 3 k$), as is the case for potentials derived in the non-supersymmetric case. In deriving the potential~\eqref{eq:Veff_J1} we have consistently dropped terms that are higher order in $|\varphi|/M_5^{3/2}$. As $\varphi_- \sim J M_5^{3/2}$ and there are no cubic terms in the Lagrangian, the corrections to~\eqref{eq:Veff_J1} will appear as $\mathcal{O}(J^4)$ terms.

In the Goldberger-Wise model a large hierarchy is generated because the stabilizing field grows slowly in the IR. For a similar kind of mechanism to work here we can choose
\begin{align}
    g_h = -g_c \left(1+\frac{2\epsilon}{3} \right) = -\frac{k(3+2\epsilon)}{2M_5} \, ,
    \label{eq:epsilon}
\end{align}
and take $\epsilon \ll 1$ so that the supersymmetric solution for $\varphi_-$ grows as $\varphi_- \sim e^{-\epsilon k y}$. In this case the effective potential becomes
\begin{align}
    V_{\rm eff} = (4 + 2\epsilon) k M_5^3 \rho^{(4 + 2\epsilon)}\left|J_0-J_\pi \rho^{-\epsilon}\right|^2 + \mathcal{O}(\rho^8)\, ,
    \label{eq:Veff_J2}
\end{align}
where we have defined the radion field of the effective theory to be 
$\rho = e^{-k\pi r}= e^{- 2 \pi g_c M_5 r/3}$.

The potential we find is a perfect square that is minimized at $V=0$. The non-supersymmetric GW model does not have the perfect square form, and is  minimized at $V=0$ only when the brane tensions are tuned to the correct values to give flat space in 4d. In our case, the tuned values of the brane tensions are guaranteed by supersymmetry (see eq.~\eqref{eq:tensions}). The potential~\eqref{eq:Veff_J2} is also smooth in the $k \to 0$ limit (keeping $g_h$ fixed by taking $\epsilon k$ constant as $k \to 0$), so is not plagued by the problems discussed in section~\ref{subsec:EFT_comparison}.\footnote{Given we have taken $e^{\frac{8}{3} \pi g_c M_5 r} \gg e^{2 \pi  M_5 r (g_c+g_h)}$ to simplify the denominator in equation~\eqref{eq:Veff_J1} this limit only makes sense if $e^{2 \pi  M_5 r g_h} \ll 1$, implying $g_h<0$.} In the next section we will present a supersymmetric 4d EFT that reproduces the above potential, which will therefore also have a sensible $k \to 0$ limit provided that it reproduces~\eqref{eq:Veff_J2}.

\subsubsection{Supersymmetric Description}

\label{subsec:SUSY_EFT}

We now write the effective potential in our model in a manifestly supersymmetric way using the 4d superspace formalism. Because we have not integrated out full superfields, we do not directly derive the effective theory. Instead we find a supersymmetric 4d EFT that reproduces the 4d potential we just derived. Before adding the SM, the 4d theory is that of the radion, which we again parameterize as $\rho = e^{-k \pi r_c}$, but now treat as a full superfield. Note that the $F$-term of the low-energy radion is not the same as that of the five-dimensional theory, and therefore has a different equation of motion. We therefore have two unknown functions to determine: the K\"ahler potential, $K(\rho, \rho^\dag)$, and the effective superpotential, $W_{\rm eff} (\rho)$. 

The K\"ahler potential is chosen to reproduce the kinetic term of the radion~\cite{Bagger:2000eh}, which comes from the Einstein-Hilbert term in 5d and is given by~\cite{Csaki:2000zn, Chacko:2013dra} 
\begin{align}
    \mathcal{L}_{\rm kin} &= \frac{3M_5^3}{k} |\partial \rho|^2 
    = 3 \MP^2 |\partial \rho|^2  \, .
\end{align}
To reproduce this term, the K\"ahler potential for the radion must be~\cite{Luty:2000ec, Gregoire:2004nn, PaccettiCorreia:2006rpo}
\begin{align}
    K_{\rm rad}(\rho, \rho^\dag) = - 3 \log \left(1 - \left | \rho \right |^2 \right) \,  .
    \label{eq:f_rad}
\end{align}
This also reproduces the correct coefficient of $\mathcal{R}^{(4)}$ that comes from integrating over the extra dimension. 

The remaining function  to determine is then the effective superpotential, $W_{\rm eff}$. This can be chosen to match the effective potential~\eqref{eq:Veff_J2} derived from the 5d model. This leads to:
\begin{align} 
    W_{\rm eff} (\rho)&= \sqrt{6(2+\epsilon) } \left(\frac{k}{M_5}\right)^{3/2}
    \left( \frac{J_0 \rho^{3+\epsilon}}{3+\epsilon} - \frac{J_\pi \rho^3}{3} \right) \, .
    \label{eq:Weff1}
\end{align}
Substituting $W_{\rm eff}$ into the expression for the potential (see eq.~\eqref{eq:eff_pot}) introduces additional terms, higher-order in $\rho$, that are not present in~\eqref{eq:Veff_J2} that we neglect. The supersymmetric description of the radion effective theory is then
\begin{align}
    \frac{ \mathcal{L}_{\rm eff}}{\sqrt{-g}} &= - 3\dfour |\C_4|^2 e^{- K_{\rm rad}/3} + \left[ \dsq \, \C_4^3 W_{\rm eff} (\rho) + h.c. \right] \, , 
    \label{eq:4d_EFT}
\end{align}
where $\C_4$, the compensator in the effective theory, is related to but not the same as the compensator, $\C_+$, in the 5d theory. 

\section{K\"ahler Potential Breaking of No-Scale Structure}

\label{sec:beta_models}

At this stage we have a 5d and 4d theory in which the radion is stabilized, but the no-scale equation of motion for $F_\Sigma$ remains. We now see what happens when the no-scale structure is broken.

In fact, the no-scale structure is naturally broken by loop corrections~\cite{Ponton:2001hq}. In five dimensions, these contribute to an extra term in the K\"ahler potential proportional to $\mathcal{V}_\Sigma^{-3}$. $\mathcal{L}_{\rm K}$ is now given by
\begin{align} \label{eq:dtermcorrection}
   \mathcal{L}_{\rm K} &= e^{2\sigma} \dfour \
   \left( \frac{\beta}{3  \mathcal{V}_\Sigma^3}  - 3 \mathcal{V}_\Sigma \right) \left[\Sigma^\dag_a \left( e^{-2g_c \sigma_3 V} \right)^a_b \Sigma^b - \Phi^\dag_a \left( e^{-2 g_h \sigma_3 V } \right)^a_b  \Phi^b \right]^{2/3}  \, ,
\end{align}
where $\beta$ is a parameter that controls the size of the correction, and should be of order $\beta \sim 1/16\pi^{2}$. As the magnitude of no-scale breaking in these models is controlled by $\beta$, we refer to these models as $\beta$-models. This form of the correction has been calculated in refs.~\cite{Rattazzi:2003rj, Gregoire:2004nn, vonGersdorff:2005bf}. The equation of motion for $F^\dag_\Sigma$ no longer relates the $F$-terms of the compensator to only the hypermultiplet $F$-terms, and can now be solved to determine $F_\Sigma$. The result is:
\begin{align}
    F_\Sigma = \frac{g_h M_5^3 r^5 }{\beta} \varphi_+^\dag \varphi_-^\dag  e^{\sigma}
    + \frac{(r \beta + 3M_5^4 r^5)}{6 \beta M_5^2} (\Fcp C_+- \Fp \varphi_+^\dag - \Fm \varphi_-^\dag ) + \mathcal{O}(\beta)\, .
\end{align}
This term also introduces an explicit kinetic term for the radion in addition to the kinetic term which comes from the Einstein-Hilbert terms. If $\beta>0$ then this contribution to the kinetic term for $r$ has the correct sign. That is the case we consider here, although in principle we could take $\beta<0$ provided that the kinetic term has the correct sign once the contribution  from $\mathcal{R}$ is included. 

\subsection{Constant Boundary Superpotentials}

\label{subsec:Wbdy}

In order to generate a compensator $F$-term we need to also include a constant superpotential, which in principle could either be in the bulk or on one of the branes. The $U(1)_R$ symmetry of the bulk prevents the inclusion of a bulk superpotential, which must come from the vacuum expectation value (vev) of some fields breaking this symmetry (we discuss this more in the following section). However, the $U(1)_R$ symmetry of the bulk theory is broken by orbifolding, so there is no symmetry obstruction preventing boundary superpotentials. Adding superpotentials to the boundaries is therefore more appealing at first glance. 

We will soon see that $W_0$ and $W_\pi$ in isolation do not lead to negative energy density or to universal anomaly-mediated masses. Instead, they shift the boundary condition satisfied by $\varphi_-$ by an amount proportional to $W_{0/\pi}\sqrt{\beta}$, with the potential still minimized at $V=0$. $W_0$ and $W_\pi$ also lead to no-scale SUSY breaking terms due to a non-zero $F_\Sigma$, with $F_\Sigma$ nonzero only on the branes. We will see that $\Fcp$ will also be nonzero on the branes, leading to anomaly-mediated mass terms for fields on the branes. Such terms do not have the universal form predicted from the 4d theory.

The boundary Lagrangian we consider is
\begin{align}
    \mathcal{L}_{\rm bdy} 
    &= 2e^{3\sigma} \left[ \delta(y) \dsq \left(\C_+^2 W_0 + 2 \C_+ J_0 \Phi_+ \right)
    - \delta(y - \pi) \dsq \left( \C_+^2 W_\pi + 2 \C_+ J_\pi \Phi_+ \right)+ h.c. \right]\, ,
    \nonumber       \\
    &= 4 e^{3\sigma} \bigg[ \delta(y) \left( C_+J_0 \Fp + \Fcp (J_0 \varphi_+ + C_+W_0) \right)+ h.c. 
    \nonumber   \\
    & \hspace{40mm}
    - \delta(y - \pi) \left( C_+J_\pi \Fp + \Fcp (J_\pi \varphi_+ + C_+W_\pi)\right) + h.c. \bigg]\, ,
    \label{eq:boundarY^2}
\end{align}
where $C_+$ and $\varphi_\pm $ have dimension $3/2$, and $W_{0/\pi}, \, J_{0/\pi}$ are dimensionless. There is no supersymmetric solution to this model, so we can't determine the boundary conditions satisfied by $\varphi_\pm$ simply by setting the $\delta$-function terms in each $F$-component to zero. Nevertheless, we do need to eliminate the badly singular terms in the potential that are proportional to  $\delta(y)^2$ and $\delta(y-\pi)^2$. We require that these vanish once the hypermultiplets satisfy their boundary conditions. Expanding to linear order in $\beta$ the dangerous terms are: 
\begin{align}
    V |_{\rm singular} 
    =&\frac{8}{M_5^4 r^5} \bigg \{ [\delta(y)]^2 \left( (M_5r)^4 \left | J_0 M_5^{3/2} -\widetilde \varphi_- \right |^2 + \beta  \left( \frac{1}{9} |J_0 M_5^{3/2} - \widetilde \varphi_-|^2 -2 |W_0|^2 \right) \right) \, , \nonumber
    \\
    &+
    [\delta(y-\pi)]^2  \left( (M_5r)^3 \left | J_\pi M_5^{3/2} 
    -\widetilde \varphi_- \right |^2
    +\beta \left( \frac{1}{9} |J_\pi M_5^{3/2} - \widetilde \varphi_-|^2 -2 |W_\pi|^2 \right) \right)\bigg \} \, ,
    \label{eq:delta-sq-terms}
\end{align}
where we have kept terms up to quadratic in combinations of the $W$'s, $J$'s and $\varphi_\pm$'s. We have also used $\varphi_- = \Theta(y) \widetilde \varphi_-$, as discussed in section~\ref{subsec:sources}. Setting $ V |_{\rm singular} =0$ then leads to the boundary conditions:
\begin{align}
    &\widetilde \varphi_- (0) = M_5^{3/2}  \left(J_0 +\frac{ \sqrt{2\beta \,} W_0}{(M_5 r)^2} \right) \, ,
    &&\widetilde \varphi_- (\pi) =  M_5^{3/2} \left(J_\pi +\frac{ \sqrt{2\beta \,} W_\pi}{(M_5 r)^2} \right)\, .
\end{align}

As promised, there is a nonzero $F$-term for both the radion and compensator, 
\begin{align}
    \Fcp^\dag \big |_{\rm bdy} &= - \frac{18 \beta e^\sigma}{M_5^{3/2} r^4} \left(\delta(y)  W_0 + \delta(y-\pi)  W_\pi \right) \, ,
    \\
    F_\Sigma^\dag \big |_{\rm bdy} &= - 2e^\sigma M_5 \left(\delta(y) W_0 + \delta(y-\pi)  W_\pi \right) + \mathcal{O} (\beta) \, ,
\end{align}
which will lead to no-scale SUSY-breaking, but only for fields with support on the boundaries (at tree level). This agrees with the results of refs.~\cite{Bagger:2001ep, Bagger:2001qi, Marti:2001iw}, which also found that constant boundary superpotentials lead to a nonzero radion $F$-term. Because these papers didn't include $\beta \neq 0$, they had vanishing $\Fcp$. There were also no $\delta$-squared terms to deal with because they didn't include hypermultiplets or sources either.

The possible SUSY-breaking mass terms coming from $\Fcp$ and $F_\Sigma$ are present only  on the same boundary to which we added the superpotential. We don't get the universal anomaly-mediated mass spectrum that we expected from a 4d analysis. In the following section we will add a superpotential in the bulk and show that this leads to anomaly-mediated masses for fields everywhere in the extra dimension, although the masses are warped so don't have a universal value. We discuss these issues in more detail in section~\ref{sec:AMSB}.

\subsubsection{Effective Potential}

The effective potential for the above model has the same form as in equation~\eqref{eq:Veff_J1} \&~\eqref{eq:Veff_J2}, after accounting for the new boundary values of $\varphi_-$. Defining $\epsilon$ as in~\eqref{eq:epsilon} we find the potential
\begin{align}
    V_{\rm eff} = (4 + 2\epsilon ) k M_5^3 \rho^{(4 + 2\epsilon)}\left| J_0 +\frac{ \sqrt{2\beta \,} W_0}{(M_5 r)^2}- \left(J_\pi +\frac{ \sqrt{2\beta \,} W_\pi}{(M_5 r)^2} \right)\rho^{-\epsilon}\right|^2 + \mathcal{O}\left(\rho^8, \beta^2 \rho^4\right)\, .
\end{align}
We can also write it in a supersymmetric way, following the approach of section~\ref{subsec:SUSY_EFT}, which leads to the effective superpotential
\begin{align}
    W_{\rm eff} (\rho)&= \sqrt{6(2+\epsilon) } \left(\frac{k}{M_5}\right)^{3/2}
    \left( \frac{\left(J_0 +\frac{ \sqrt{2\beta \,} W_0}{(M_5 r)^2}\right) \rho^{3+\epsilon}}{3+\epsilon} - \frac{\left(J_\pi +\frac{ \sqrt{2\beta \,} W_\pi}{(M_5 r)^2}\right)\rho^3}{3} \right) \, .
    \label{eq:Weff2}
\end{align}
Any SUSY breaking masses that are generated by $F_\Sigma$ and $\Fcp$ in this model must be included in the matching as explicit SUSY-breaking soft terms. 

A notable consequence is that if there is a gauge field in the bulk with even parity the radion $F$-term will generate a SUSY-breaking mass for the gauginos, which should be included as an explicit SUSY-breaking term in the EFT. The gauginos can also communicate SUSY-breaking to fields on the boundaries, which can be the dominant effect for fields charged under the new gauge group, and can help to resolve the problem of wrong-sign slepton masses~\cite{Chacko:2000fn}.

\subsection{Boundary Superpotentials as Sources}

\label{subsec:Wbdy-only}

The analysis of the previous section shows that boundary superpotentials do not trivially integrate to constant superpotentials in the 4d theory. Instead they lead to SUSY breaking terms due to a nonzero $F_\Sigma$ and act as sources for the bulk fields when $\beta \neq 0$. As written, this model does not source negative (AdS$_4$) energy density or a universal anomaly-mediated mass spectrum. In the next section we will see the necessity of a bulk superpotential to construct a realistic model of anomaly mediation.

But before presenting the bulk superpotential, we first  note the interesting case where $J_0 = J_\pi=0$. Here  $\varphi_-$ is sourced only by $W_{0/\pi}$ and the effective potential is 
\begin{align}
    V_{\rm eff} = \frac{ 4k \beta (2 + \epsilon ) }{M_5 r^4} \rho^{(4 + 2\epsilon)}\left| W_0 - W_\pi \rho^{-\epsilon}\right|^2 + \mathcal{O}\left(\rho^8, \beta^2 \rho^4\right)\, .
\end{align}
Notice that neither $W_0$ nor $W_\pi$ appear as constants in $W_{\rm eff}$, but instead multiply powers of $\rho$.  $W_0, W_\pi$ act as sources for the bulk hypermultiplets when the no-scale structure is broken, so appear with the $r$-dependence of the bulk $\varphi_\pm$ fields obtained from solving the bulk field equations. The effective superpotential comes from taking the $J \to 0$ limit of eq.~\eqref{eq:Weff2}:
\begin{align}
    W_{\rm eff} (\rho)&= 2 \sqrt{3\beta (2+\epsilon) } \frac{k^{3/2}}{M_5^{7/2} r^2}
    \left( \frac{W_0 \rho^{3+\epsilon}}{3+\epsilon} - \frac{ W_\pi\rho^3}{3} \right)
\end{align}

We can contrast this to the effective superpotential we would have obtained if we had integrated over $y$ before deriving the potential in the 5d model:
\begin{align}
    W_{\text{na\"ive 4d}} (\rho)&\sim  \left(\frac{k}{M_5}\right)^{3/2}
    \left(  W_0 - W_\pi \rho^3  \right) \, ,
    \label{eq:W_naive}
\end{align}
which has been assumed in the past. This would lead to an effective potential
\begin{align}
    V_{\text{na\"ive 4d}} (\rho)&= 3\MP^4\left( \rho^4 |W_\pi|^2 - |W_0|^2 \right)
\end{align}
As already discussed, the powers of $\rho$ that $W_{0/\pi}$ multiply in the correct EFT are different because they act as sources for the bulk fields. We see that the true answer for $W_{\rm eff}$ is  only non-zero when the bulk no-scale structure is broken. This is not evident within the 4d EFT alone, because $\rho$ doesn't have a no-scale kinetic term (see eq.~\eqref{eq:f_rad}) even though the no-scale structure is present in the bulk.

The (incorrect) effective superpotential in equation~\eqref{eq:W_naive} has been used in models looking to generate a hierarchy of scales between the gravitino mass and anomaly-mediated masses--in other words models where $F_{C_4} \ll m_{3/2}$ (see, e.g.~\cite{Marti:2001iw,Luty:2002ff,Arkani-Hamed:2004zhs}). In models without warping or a no-scale structure, $m_{3/2} \propto W$ and $F_{C_4} \propto W$, so this hierarchy is not present. The hope in warped models was for a separation of scales because the superpotentials were on different branes, with the IR brane superpotential being suppressed due to the warping.

In future sections we will see that all anomaly-mediated masses, whether they originate from a bulk or boundary superpotential, come with a warp factor. However, because a nonzero $\Fcp$ requires a breaking of no-scale structure, the anomaly-mediated masses are further suppressed relative to $m_{3/2}$ by the no-scale breaking parameter $\beta$. These factors and contributions from both bulk and boundary superpotentials tell us that  despite the apparent universal nature of anomaly-mediated masses, a wide variety of mass hierarchies is in principle possible. As an aside, we  note that  a hierarchy between scalar and gaugino masses is natural for masses not arising from anomaly-mediation, as has been suggested for split supersymmetric models, for example~\cite{Giudice:2004tc}.

\subsection{An AdS$_4$ Solution From K\"ahler Potential No-Scale Breaking}

\label{subsec:AdS_loop}

So far, all models are minimized at $V=0$, with the compensator $\Fcp = 0$ or nonzero only on the boundary. This does not give a phenomenologically viable model of anomaly mediation because any SUSY-breaking terms on the branes will lead to a dS$_4$ vacuum.\footnote{Despite the positive cosmological constant today, the value is so small that to match our Universe we need to cancel any supersymmetry breaking energy with a negative contribution. The latter is the challenge in the set-up so far.}

We need a source of negative energy density to address this issue. We have seen in the previous section that a boundary superpotential does not in itself achieve this. In this section we include a constant bulk superpotential, $W_{\rm bulk}$. The gauged $U(1)_R$ symmetry forbids such a bulk term unless the symmetry is broken, so $W_{\rm bulk}$ must  be sourced by expectation values of fields with $U(1)_R$ charge. One such example  is fluxes from higher-form fields in a higher dimensional space if we imagine our 5d theory as an effective theory descending from a 10d string compactification~\cite{Gukov:1999ya}. Alternatively, $W_{\rm bulk}$ could come from gaugino condensation. For our purposes, we remain agnostic to the source of $W$ and include it in the bulk by adding the term:
\begin{align}
    \Delta \mathcal{L}_{\rm flux} = - M_5 r \, e^{3\sigma} \left[ \dsq W_{\rm bulk} \left( \C_+^2 + \C_-^2\right) + h.c.  \right] \, ,
    \label{eq:W_const}
\end{align}
with the understanding that this term likely has an origin in gauge fields or fluxes. The factor of $M_5 r$ is necessary to reproduce the correct $r$-dependence of the 5d potential. This is still insufficient to generate a flat, SUSY-breaking minimum  if the no-scale form of the K\"ahler potential remains unbroken. Here we extend the analysis of the  previous section and include the terms in eq.~\eqref{eq:dtermcorrection} that break the no-scale structure by loop corrections. 

Neglecting $\varphi_\pm$, which we assume are suppressed, the $F$-terms for $\Sigma$ and the compensator are given by
\begin{align}
    F_\Sigma &= - M_5^2 r W_{\rm bulk}\, e^{\sigma (y)}   + \mathcal{O}(\beta) 
    \, , \nonumber    \\
    \bFcp &= -  \frac{ 2 \beta \, W_{\rm bulk}}{M_5^{3/2} r^4} \, e^{\sigma (y)} + \mathcal{O}(\beta^2) \, .
    \label{eq:FC_sol1}
\end{align}
The leading correction to the potential from the bulk superpotential is then:
\begin{align}
    \Delta V_\eff = - \frac{M_5 \beta}{k r^4} \left(1 - e^{-4k\pi r_c} \right)  \, |W_{\rm bulk}|^2 + \mathcal{O}(\beta^2)\, .
    \label{eq:DeltaV_beta}
\end{align}
We treat this as a perturbation to the potential~\eqref{eq:Veff_J1} generated by sourcing the hypermultiplets on the boundaries, which assumes we are working in the regime
\begin{align}
    \frac{\sqrt{\beta}}{M_5 k r^2} |W_{\rm bulk}| \ll  |J_{0/\pi}| e^{-2k\pi r_c} \, .
    \label{eq:inequality}
\end{align}

The term~\eqref{eq:DeltaV_beta} is the Casimir energy contribution~\cite{Ponton:2001hq}, and is negative for $\beta >0$ due to the constant $\Fcp$ in the bulk, generating an AdS$_4$ minimum. Here we see that when $\beta = 0$ we recover the no-scale SUSY-breaking model, where the nonzero $F_\Sigma$ breaks supersymmetry but the potential vanishes at the minimum. This has been well studied in models with flat extra dimensions~\cite{Marti:2001iw, Bagger:2001ep,Antoniadis:1998sd}. Here we show that the same conclusions holds even with warping, which was previously overlooked because the calculations were done in the 4d EFT. The energy at the minimum comes predominantly from the $F_\Sigma \Fcp$ terms, but can be written in terms of $\Fcp$ as
\begin{align}
    \Delta V_\eff \simeq - \frac{3(M_5 r)^4}{4 \beta k} \, |e^{-\sigma}\Fcp|^2 \, ,
\end{align}
so for $\beta \ll (M_5 r)^4$ we have $\Delta V_\eff \gg e^{-\sigma}\Fcp$. Note that this means that anomaly-mediated masses would be smaller than the typical expectation based on the negative energy contribution from the potential.

The AdS minimum can be reproduced in the 4d EFT by adding the term
\begin{align}
    \delta W_{\rm eff} =\frac{\sqrt{\beta k} W_{\rm bulk}}{\sqrt{3} M_5^{5/2} r^2} \, ,
\end{align}
to the superpotential~\eqref{eq:Weff1}. The effective superpotential for the model with constant superpotential in the bulk and vanishing boundary superpotentials is then:
\begin{align} 
    W_{\rm eff} (\rho)&= \frac{\sqrt{\beta k} W_{\rm bulk}}{\sqrt{3} M_5^{5/2} r^2} +  \sqrt{6(2+\epsilon) } \left(\frac{k}{M_5}\right)^{3/2}
    \left( \frac{ J_0 \rho^{3+\epsilon}}{3+\epsilon} - \frac{ J_\pi \rho^3}{3} \right) \, .
    \label{eq:Weff3}
\end{align}

This reproduces the leading $\rho$-independent AdS$_4$ energy term of ~\eqref{eq:DeltaV_beta} and the stabilizing potential~\eqref{eq:Veff_J1}. However, it also leads to spurious terms in the potential proportional to $W_{\rm bulk}$, $J_{0/\pi}$, and powers of $\rho$. These can be cancelled with  extra terms in $W_{\rm eff}$, but we simply neglect these terms as they are subleading. As discussed previously, there are also SUSY-breaking soft terms that must be included in the low-energy potential due to $F_\Sigma \neq 0$, that can't be reproduced in a supersymmetric EFT. 

\section{Superpotential No-Scale Breaking}

\label{sec:condensates}

As we have already described, generating a negative energy density requires a bulk superpotential as well as breaking the no-scale form of the K\"ahler potential. Negative energy requires breaking the gauged $U(1)_R$ symmetry, which, when preserved, forbids adding constant terms to the bulk superpotential. This means that a constant bulk superpotential must arise from the expectation values of fields that break the symmetry. 

So far we have considered breaking of the no-scale form of the K\"ahler potential by loop corrections.
An alternative way to break the no-scale structure could be gauginos in the bulk that condense, which ultimately breaks $U(1)_R$ and can obviate the need for an independent stabilization sector.
Such models have been considered as models to stabilize both flat and warped extra dimensions~\cite{Luty:1999cz, Luty:2000ec}, and are realized in many string theory setups~\cite{Horava:1996vs, Lalak:1997zu, Nilles:1997cm, Antoniadis:1997xk, Lukas:1998tt, Lukas:1998yy, Kachru:2003aw, Hamada:2018qef, Kallosh:2019oxv, Kachru:2019dvo, Hamada:2021ryq}. 
The gauge coupling in the 4d theory, and therefore the condensate vev, depends on $\Sigma$. This leads to  an additional term in the $F_\Sigma$ equation of motion that  breaks the no-scale structure and introduces $r$-dependence in the effective potential that can lead to a stabilized model~\cite{Burgess:1995aa}.

For this class of models we include a non-abelian gauge group in the bulk and  set the bulk hypermultiplets to zero, $\varphi_\pm = 0$. We include  a vector superfield $A$ with field strength $G_\alpha$ through the terms\footnote{There is also a chiral superfield, $S$, associated with the gauge field. Now that we have multiple gauge sectors in the bulk there is also the possibility of having a more complicated prepotential, which would lead to $S$-dependent terms in the coefficient of $G^{\alpha} G_\alpha$ and terms coupling the components of $V$ and $A$~\cite{Seiberg:1996bd}. It would also alter the definition of the radion, so in order not to deviate too far from the previous discussion we do not consider that possibility here, so $S$ will play no role in our analysis.}
\begin{align}
    \Delta \mathcal{L}_{\rm vec} &=  - \left[ \dsq \frac{3}{2} \left\{ \Sigma \,  G^{\alpha} G_\alpha - \frac{1}{12} \overline D^2 \left( A D^\alpha \partial_y A - D^\alpha A \partial_y A \right) G_\alpha \right\} + \text{h.c.} \right] \, .
    \label{eq:gauginos}
\end{align}
In a 5d theory the running of the gauge coupling follows a power law, so the gauginos will not condense and it is generally assumed the gauge coupling takes some fixed point value. Below the compactification scale the theory becomes effectively four-dimensional, as only the zero modes survive, and the gauge coupling runs logarithmically. One would of course like to derive the answer in the 4d theory, but we do the analysis in 5d to ensure  the correctness of the  4d theory, which as we will see, particularly in the warped case, is not the na\"ive version of the theory. 

For an asymptotically free theory the low-energy running allows for gaugino condensation. As the effective 4d gauge coupling is proportional to $\Sigma$, the condensate will generate the $\Sigma$-dependent terms needed to break the no-scale structure. The Lagrangian  associated with the condensate is~\cite{Burgess:1995aa}
\begin{align}
    \Delta \mathcal{L}_{\lambda} &=
    - e^{3\sigma} \left[ \dsq \, \left( \C_+^2 + \C_-^2 \right) W_\lambda e^{- 2 \alpha \Sigma}+ h.c. \right] \, ,
\end{align}
where $W_\lambda \propto \langle \lambda \lambda \rangle/ M_5^3 $, and $\alpha$ is determined by the $\beta$-function for the coupling~\cite{Burgess:1995aa}. This term breaks the $U(1)_R$ symmetry of the bulk. The terms in equation~\eqref{eq:gauginos} are invariant under $U(1)_R$, and the breaking originates from $ \langle \Sigma \rangle \neq 0$. 

\subsection{Bulk Superpotential}

The gaugino condensate by itself is insufficient to stabilize the model, however.  To get a fully consistent model we also include a constant superpotential. In this section we add the superpotential in the bulk via the term~\eqref{eq:W_const},  $\Fcp$ and $F_\Sigma$ are then given by
\begin{align}
    \Fcp^\dag &= \alpha e^{\sigma (y) - M_5 r\alpha} M_5^{5/2} W_{\lambda} \, , \nonumber
    \\
    F_\Sigma^\dag &= - \frac{M_5 e^{\sigma (y)}}{3} \left(3M_5 r W_{\rm bulk} + e^{-M_5 r \alpha} W_{\lambda} (3 + 2 M_5 r \alpha)\right) \, .
    \label{eq:gc_Fs}
\end{align}
$\Fcp$ scales only with $W_\lambda$ and not with  $W_{\rm bulk}$, which is expected as it must vanish in the limit $W_\lambda \to 0$ where the no-scale form is restored.

After substituting the solutions for the $F$-components and integrating over $y$, the potential for this model is
\begin{align}
    V^{(1)}_{\rm gc} &= \frac{\alpha M_5^4}{6 k r}  \left(1 - e^{-4k\pi r} \right) e^{- M_5 r \alpha}  
    \left( 3M_5 r W^\dag_{\rm bulk}W_\lambda + |W_\lambda|^2 (3+\alpha M_5 r)e^{-M_5 r \alpha} + h.c. \right) \, , 
    \label{eq:Veff_gc1}
\end{align}
where $V^{(1)}_{\rm gc}$ refers to the model with $W_{\rm bulk}$ and we will use $V^{(2)}_{\rm gc}$ to refer to the potential in a model with a boundary superpotential in eq.~\eqref{eq:Veff_gc2}.
Here we see once again that $\Fcp$ is proportional to the breaking of the no-scale form, which in this case is controlled by $W_\lambda$.  There can also be corrections to the K\"ahler potential from the gaugino condensate itself which would generate contributions to $\Fcp$ proportional to $W_{\bulk}$ multiplied by the coefficient of this term, analogous to the terms in equation~\eqref{eq:FC_sol1}. These terms are subleading, however, and don't change our overall conclusions so we neglect them.

In the limit where we can drop the $e^{-4k\pi r}$ term the potential is minimized for
\begin{align}
     2|W_\lambda|^2 e^{- M_5 r \alpha} (3+ 6 M_5 r\alpha + 2(M_5 r\alpha)^2)  + 3 (M_5 r\alpha)^2 \left( W_\lambda^\dag W_{\rm bulk} + W_\lambda W^\dag_{\rm bulk} \right)= 0 \, .
\end{align}
If $|W_\lambda| \gg |W_{\rm bulk}|$ and the two superpotentials take opposite sign (so that $r$ is stabilized when ${\alpha M_5 r \gg 1}$), a solution can be found where
\begin{align}
     e^{- M_5 r \alpha} \simeq - \frac{3 \left( W_\lambda^\dag W_{\rm bulk} + W_\lambda W^\dag_{\rm bulk} \right)}{ 4 \alpha  |W_\lambda|^2}  \, .
     \label{eq:rmin_gc}
\end{align}
After substituting the solution~\eqref{eq:rmin_gc}, $F_\Sigma$ vanishes provided that the combination $W_\lambda W_{\rm bulk}^\dag$ is real, meaning that SUSY is unbroken at tree level. The value of the potential at the minimum is then
\begin{align}
    V^{(1)}_{\rm gc} |_{\min} & = -\Lambda^4_\ads \simeq - \frac{ \alpha^2  M_5^5}{3 k}  e^{- 2M_5 r \alpha} |W_\lambda|^2  \, ,
\end{align}
leading to an AdS$_4$ solution with AdS$_4$ energy scale given by $\Lambda_\ads$. This can be written in terms of the vev of $\Fcp$ as 
\begin{align}
    \Lambda^4_\ads& \simeq 
    \frac{\left | \langle e^{- \sigma} \Fcp \rangle \right |^2}{3k}  \, .
\end{align}

$F_\Sigma$ and the hypermultiplet $F$-terms vanish in this model, but supersymmetry is broken as the graviphoton $D$-term, eq.~\eqref{eq:warp_SUSY}, is nonzero. This is due to the backreaction on the metric by the 4d cosmological constant, as $D\propto e^{2\sigma} \left( \dot \sigma -k\right) \sim \Lambda^4_\ads /\MP^2 $. $D$ is constant across the fifth dimension, and can generate slepton masses for hypermultiplets charged under the $U(1)$ gauged by the graviphoton. $\Fcm$ will also be nonzero, as it is also proportional to the combination~$ \dot \sigma -k$. Notice these terms are subleading in AdS$_4$ space as they scale as the energy density, not the square root of the energy density.

The 4d AdS$_4$ space is generated as the bulk AdS$_5$ energy gets an extra contribution from the stabilizing sector, so that the boundary tensions are now below their critical values. This mismatch of tensions is what leads to an AdS$_4$ solution and SUSY-breaking due to the nonzero $D$-term. Recall that for nonzero cosmological constant, the required detuning of the IR boundary tension is enhanced by an inverse warp factor over that of the UV brane~\cite{DeWolfe:1999cp, Karch:2000ct}. The boundary tensions in our model are equal in magnitude so do not reflect this. For a fully consistent model, the boundary tensions must adjust so that the boundary conditions are satisfied by $\sigma$ in AdS$_4$ slicing~\cite{Lust:2025vyz}. This could happen because of the backreaction from a scalar field in the gauge multiplet, for example, or from boundary terms involving gauginos or scalar fields. This guarantees consistency of the AdS$_4$ theory, but won't affect the supersymmetry breaking masses at leading order. In any case, in Minkowski space, these terms are no longer relevant. We also note that this differs from ref.~\cite{Bagger:2003dy}, which found a supersymmetric solution with brane tensions detuned in the correct way to generate a stable AdS$_4$ solution without a stabilizing sector.

\subsection{Boundary Superpotentials}

We now ask what happens if the constant superpotential is on the boundary as opposed to in the bulk, including $W_0$ as in equation~\eqref{eq:boundarY^2} (with the sources $J_0, J_\pi$ set to zero).\footnote{We could include an IR superpotential $W_\pi$, but terms proportional to $W_\pi$ are suppressed by additional powers of the warp factor, so are only significant when $W_0 = W_{\rm bulk} = 0$.} In this case, the radion $F$-component is non-vanishing and is 
\begin{align}
    &F_\Sigma^\dag = - \frac{M_5 e^{\sigma (y)}}{3} \left[ W_{\lambda} e^{- M_5 r \alpha} \left(3 + 2 M_5 r \alpha \right) + 6W_0\delta(y)  \right] \, ,
\end{align}
whereas the compensator $\Fcp$ is the same as the model with a bulk superpotential, and is given in~\eqref{eq:gc_Fs}. 

The  effective potential is now
\begin{align}
    V^{(2)}_{\rm gc} &= \frac{\alpha M_5^4}{3 k r}  e^{- M_5 r \alpha} 
    \left(6 k r W_\lambda^\dag  W_0  
     + \frac12 (1 - e^{-4 k \pi r} ) (3+\alpha M_5 r)  |W_\lambda|^2 e^{- M_5 r \alpha} + h.c. \right) \, .
    \label{eq:Veff_gc2}
\end{align}
In the limit where we can drop $\mathcal{O}\left( e^{-4k \pi r} \right)$ terms, this leads to
\begin{align}
    V^{(2)}_{\rm gc} &= \frac{\alpha M_5^4}{3 k r}  e^{- M_5 r \alpha} 
    \left(6kr W_\lambda^\dag W_0 + \frac12 (3+\alpha M_5 r) |W_\lambda|^2 e^{- M_5 r \alpha}+ h.c. \right) \, .
    \label{eq:Veff_gc3}
\end{align}
Again we find a consistent solution where $\alpha M_5 r \gg 1$, and find the same form for the potential at the minimum as when the constant superpotential originates in the bulk:
\begin{align}
    V^{(2)}_{\rm gc}|_{\min} &= -\frac{\alpha^2 M_5^5  e^{- M_5 r \alpha} |W_\lambda|^2 }{3k} 
    =  -\frac{\left | \langle e^{- \sigma} \Fcp \rangle \right |^2 }{3k} \, .
    \label{eq:ads_min}
\end{align}

With the no-scale form broken by the gaugino condensate, boundary superpotentials do allow for supersymmetry breaking from the radion and an AdS$_4$ minimum that allows for  anomaly mediation.  However, despite the non-zero $F_\Sigma$, the integral of $F_\Sigma$ vanishes (to leading order in the warp factors):\footnote{This assumes there is no relative phase between $W_0$ and $W_\lambda$.}
\begin{align}
    \int_0^\pi dy F_\Sigma = 0 \, ,
\end{align}
so $F_\Sigma$ does not generate masses for fields with flat wavefunctions in the bulk, such as gauginos. If $F_\Sigma$ is integrated against other fields with non-trivial bulk profiles, it can lead to SUSY-breaking masses. 
If there are loop corrections to the K\"ahler potential in the form of a $\beta$-term (see eq.~\eqref{eq:dtermcorrection}) then the boundary superpotentials would also act as sources for $\varphi_-$, as discussed in section~\ref{subsec:Wbdy-only}.

\subsection{Comparison to Na\"ive EFT}

\label{subsec:gc_comparison}

In section~\ref{sec:beta_models} we found that the superpotential of the 4d theory was not simply the integral of the 5d superpotential in the $\beta$-models. There were two essential sources for this discrepancy. First, in warped geometry, the K\"ahler potential in the EFT does not reflect the no-scale structure of the bulk theory. Second, when the scalar fields $\varphi_\pm$ have non-trivial profiles in the bulk, they lead to $\rho$-dependent terms to the effective superpotential. Based on these observations, we might expect agreement with the na\"ive effective theory with a flat extra dimension, unbroken supersymmetry, and constant bulk fields. We now show this is indeed the case.

In this section we compare the potentials obtained above to the na\"ive superpotential $W_{\text{na\"ive 4d}}$ generated by integrating over $y$ before solving for the auxiliary fields. This leads to:
\begin{align}
     W_{\text{na\"ive 4d}} &=  W_0 + W_\pi e^{-3 k \pi r}  +  \frac{\left(1 - e^{-3 k \pi r} \right) }{3 k r} \left( M_5 r W_{\rm bulk}  + W_\lambda e^{-2\alpha \Sigma} \right) \, ,
    \label{eq:W_naive4d}
\end{align}
where we have ignored any contributions from the hypermultiplets and kept the $M_5 r$ factor which was included to get the correct $r$-dependence in 5d. In the previous section we dropped terms subleading in $e^{-k \pi r}$ in the potential and set $W_\pi = 0$. To match this solution we again set $W_\pi =0$ and take the $k \to 0$ limit, where the superpotential reduces to
\begin{align}
     W_{\text{na\"ive 4d}} & \to W_0 + \pi M_5 r W_{\rm bulk} + \pi W_\lambda e^{-2 \alpha \Sigma}  \, .
     \label{eq:W_naive4d_gc}
\end{align}
We will compare the potential one obtains in the EFT with the above superpotential to the leading terms of the full answer from the 5d theory. When we include both a constant superpotential both in the bulk and on the UV boundary, the full potential we want to reproduce is given by:
\begin{align}
    V_{\rm gc, \, 5d} &= \frac{\alpha M_5^4}{6kr}   e^{- M_5 r \alpha} 
    \left( 3r W_\lambda \left(4 k W_0^\dagger + M_5 W^\dag_{\rm bulk} \right) + |W_\lambda|^2 (3+\alpha M_5 r)e^{-M_5 r \alpha} + h.c. \right) \, ,
    \label{eq:V5d}
\end{align}
where we again have dropped $\mathcal{O}(e^{-4k\pi r})$ terms.

\subsubsection{Warped Extra Dimension}

The  K\"ahler potential for the warped effective theory, given in eq.~\eqref{eq:f_rad}, is:
\begin{align}
    \mathcal{L}_{K, \text{warped EFT}} = -3 \dfour \left( 1 - \left | \rho \right |^2 \right)  |\C_+|^{4/3} \, .
    \label{eq:LK_warpedEFT}
\end{align}
This K\"ahler potential with the superpotential $W_{\text{na\"ive 4d}}$ in eq.~\eqref{eq:W_naive4d_gc} was the model considered, for example, in ref.~\cite{Luty:2000ec}. The potential in this model would be:
\begin{align}
    V_{\text{na\"ive 4d}} &= 
    \frac{ M_5^6}{3k^2 }
    \left( 4\pi^2 |W_\lambda|^2 e^{2(\pi k r - 2\alpha  M_5r)} -\left| 3\left( W_0 + \pi M_5 r W_{\rm bulk} \right) + W_\lambda e^{-2\alpha  M_5 r} \left(1 + \frac{2\alpha M_5 r}{3k} \right) \right |^2 \right) + \mathcal{O}(\rho^2) \, .
    \label{eq:warped_Vgc}
\end{align}
This does not reproduce the effective potentials of equations~\eqref{eq:Veff_gc1} and~\eqref{eq:Veff_gc2}, even to leading order in $e^{-k\pi r}$. In fact we see that the first term above is exponentially large in the regime $k \pi > \alpha M_5$.\footnote{In ref.~\cite{Luty:2000ec} the authors work in the opposite limit so were able to generate a stabilized hierarchy.} There are also terms of order $e^{-4 \alpha  M_5 r}$ which are not present in $V_{\rm gc, \, 5d}$. $V_{\text{na\"ive 4d}}$ also diverges in the $k \to 0$ limit, while all the potentials derived from the 5d theory have a smooth $k \to 0$ limit. This is related to the divergence in the $k\to0$ limit we found for the toy model in section~\ref{subsec:EFT_comparison}.

Here we see explicitly that simply integrating over the superpotential in the 5d theory to derive the 4d superpotential does not reproduce the full 5d analysis in warped compactifications. To derive the correct EFT, as we did in section~\ref{subsec:EFT}, we can construct an effective superpotential by requiring that the 4d supergravity potential reproduces the potential of the 5d theory. Given we know the K\"ahler potential, the task is to find a $W_{\rm eff, \, gc }  (\rho)$ which reproduces $V_{\rm gc, \, 5d}$. Defining $x = \alpha M_5/(k \pi)$ so that $\rho^{x} = e^{- M_5 r \alpha}$, the leading terms in the effective superpotential are given by
\begin{align}
    W_{\rm eff, \, gc }  (\rho)&= 
   \frac{1}{4M_5}\sqrt{\frac{\alpha  k r}{3 + \alpha M_5 r}} \left[ \left(4 k W_{0}+M_5 W_{\rm bulk} \right)(1 + 3 \rho )
    + \frac{4 (3 + \alpha  M_5 r) }{r(1+x)} W_\lambda \rho^{1+x} \right]\, .
    \label{eq:Weff_gc}
\end{align}
This superpotential reproduces only the leading terms in~\eqref{eq:V5d}, with the first correction coming at $\mathcal{O}(e^{-k\pi r})$. This can be rectified by adding higher powers of $\rho$ to the functions multiplying $W_0$ and $W_\lambda$ in equation~\eqref{eq:Weff_gc} so that the potential obtained from $W_{\rm eff, \, gc }$ matches the subleading terms, but we do not do this here. In deriving~\eqref{eq:Weff_gc} from the general formula for the supergravity potential (eq.~\eqref{eq:eff_pot}), we have used $\MP^2 = M_5^3/k$.

\subsubsection{Flat Extra Dimension}

The K\"ahler potential for the effective theory in the limit of flat extra dimensions also has the no-scale form 
\begin{align}
    \mathcal{L}_{K, \text{flat EFT}} = -3 \dfour \left(\Sigma + \Sigma^\dag \right) |\C_+|^{4/3} \, .
\end{align}
This can be obtained from the 5d theory by integrating over $y$ at the level of the K\"ahler potential, but is also known from an EFT analysis (see, e.g.~\cite{Luty:1999cz}). The potential we find in this case is
\begin{align}
    V_{\text{na\"ive 4d}} &= \frac{2\pi \alpha M_5^4}{3} e^{-\alpha  M_5 r} \left( 3W_\lambda \left( W_0^\dagger + \pi M_5 r W^\dag_{\rm bulk} \right) + \pi |W_\lambda|^2 (3+\alpha M_5 r)e^{-M_5 r \alpha} + h.c. \right) \, ,
    \label{eq:V_naive4d_flat}
\end{align}
which has the same form as the potential derived from the bulk theory, but with different $\mathcal{O}(1)$ coefficients that can be absorbed into the definition of the 4d superpotentials. To obtain~\eqref{eq:V_naive4d_flat} from the formula eq.~\eqref{eq:eff_pot} in this case we use the flat-space relation $\MP^2 = M_5^3r$.
\\

When the extra dimension is flat, the na\"ive integration works (up to $\mathcal{O}(1)$ factors that can be absorbed in a rescaling the superpotentials) because all quantities are constant throughout the bulk. The warped case does not, however, both because the potential terms have non-trivial bulk profiles and because the K\"ahler potential for the EFT in the warped case, eq.~\eqref{eq:f_rad}, is blind to the fact that the 5d theory had a no-scale structure. The flat case works because the K\"ahler potential of the 4d theory has the same no-scale structure as the bulk theory, SUSY was unbroken in 5d (up to possible boundary terms from $F_\Sigma$ which need to be added separately), and because the energy density was constant throughout the bulk so the integration over $y$ is trivial after appropriate redefinitions to remove factors of $\pi$. This is not the case in the effective theory of the warped extra dimension, and performing the same procedure of directly integrating the superpotential leads to exponentially enhanced terms not present in the 5d theory (see eq.~\eqref{eq:warped_Vgc}).

\subsection{Comment on Boundary Condensates \& KKLT}

\label{subsec:KKLTsingular}

A notable example of a model which is stabilized by an interplay between a constant superpotential and a gaugino condensate is the KKLT model~\cite{Kachru:2003aw}. There were several controversial aspects of this proposal but notable among them was the cancellation of singularities. In ref.~\cite{Kachru:2019dvo} a full 10d analysis was done, but ref.~\cite{Hamada:2021ryq} argued it involved nonlocal counterterms. In ref.~\cite{Hamada:2021ryq}, there was an attempt to cancel the singularities with local counterterms but as we argued in section~\ref{sec:5d_issues} this will lead to a nonrenormalizable theory. However, shifting away the singularity, as was done in \cite{Hamada:2018qef, Kallosh:2019oxv, Hamada:2021ryq}, sources a bulk field, much as we have found when we had potentially dangerous singularities, and should be a suitable way of dealing with singularities. However, the task of evaluating the finite terms is still left as an open question.

In this section we comment on an analog of the KKLT model in our 5d setup. In this simple analog model we find that the potential  has  singular terms that do not cancel. However, this is likely an artifact of the five-dimensional theory  we use here. Nonetheless, our methodology could prove useful for resolving this issue in the full theory and for illustrating some relevant issues. We comment on how this may be accomplished in a more complete model, but leave the details for future work.

The KKLT setup consists of a stack of D7 branes wrapping a 4-cycle in the Calabi-Yau, there are then two further dimensions orthogonal to the branes. If the volume of the 4-cycle is small then we can integrate over those 4 dimensions, leaving a 6d theory with a stack of 4-dimensional branes on the boundary. Even if the dimensions are not small, the superpotential from gaugino condensation will be present on the boundary, similar to the assumed 5d gaugino condensate that is generated only below the compactification scale. We are taking our 5-dimensional setup with branes at the endpoint as an approximation to this theory. That is, we replace the two extra compact dimensions by our one compact dimension, and take our 4d branes to represent the D7 branes wrapped on the 4-cycle. We emphasize this is simply to model the condensate on a singular space and has nothing to do with the anti-brane or the conifold.

In this model there are two moduli: the radion $\Sigma$ in the bulk that we have been considering, and an additional modulus, $\Sigma'$, which sets the 4-cycle volume. We make the same assumptions as in the KKLT model, namely that $\Sigma'$ has a no-scale K\"ahler potential and the volume modulus ($\Sigma$ in our model) is stabilized by fluxes and can be integrated out. The superpotential in the KKLT model consists of a constant $W_{\rm bulk}$ in the bulk and a gaugino condensate on the boundary, where the gauge coupling for the condensate is set by $\Sigma'$.

Sticking to a flat extra dimension for simplicity, our model is
\begin{align}
    \frac{\mathcal{L}_{\rm KKLT} }{\sqrt{-g}} =& 
    -3 \dfour (\Sigma'+\Sigma'^\dag) \left( |\C_+|^2 +  |\C_-|^2 \right)^{2/3}  
    - \left[ \dsq \left( rM_5 W_{\rm bulk} \left( \C_+^2 + \C_-^2 \right) + 2 \delta(y) \C_+^2 W_\lambda e^{-2\alpha \Sigma'} \right) +  h.c. \right]
\end{align}
Notice that even though the potential for $\Sigma'$ is on the boundary, $\Sigma'$ is a bulk field as it sets the Planck scale. The $F$-terms in the model are:
\begin{align}
    &\Fcp^\dag = 2 \delta(y) \alpha M_5^{5/2} W_\lambda e^{-v'\alpha} \, ,
    &F_\Sigma'^\dag = -M_5^2 r W_{\rm bulk} +  (6+4 v' \alpha)  \delta(y) \frac{W_\lambda M_5}{3} e^{-\alpha v'} \, ,
\end{align}
where $\Sigma' = v' + \theta^2 F_\Sigma'$ (ignoring the fermionic components), where $v'$ is the dimensionless volume of the 4-cycle on the boundary. The potential we find is
\begin{align}
    V_{\rm KKLT} &= 4 \alpha M_5^5 r \left( W_\lambda W_{\rm bulk}^\dag e^{-v'\alpha} + h.c.\right) + \frac{16 \alpha}{3} (3 + v' \alpha) \left | W_\lambda e^{-\alpha v'} \delta(y) \right|^2 \, .
\end{align}
We see that in our toy model for KKLT there are singular terms proportional to $\delta(y)^2$. These terms can't be cancelled by turning on hypermultiplets, as the $\delta(y)^2$ potential terms coming from the hypermultiplets are strictly positive (see eq.~\eqref{eq:delta-sq-terms}). These terms were also found in refs.~\cite{Kachru:2019dvo, Hamada:2021ryq} which argued that there should be counterterms to remove the singular pieces, while leaving a finite $|W_\lambda|^2$ term. However, the consistency of such a theory and the calculation of finite terms is unclear. Another possibility is that the singular terms act as a source for bulk fields that we have not included in our model. In six dimensions a natural candidate for this could be the two-form. If this were the case this would lead to a cancellation of singular terms through a ``perfect square" form of the potential (see e.g.~\cite{Horava:1996vs, Kallosh:2019oxv}).

We leave these as open questions. With the KKLT model assumptions, all the ingredients necessary to stabilize the radion and generate AdS$_4$ space are in principle present. Based on our effective theory analyses described in the following section, we expect that if the two extra dimensions are flat, the 4d theory will indeed agree with the higher-dimensional approach. However, we have also seen that modulus stabilization can break supersymmetry when gravitational corrections to the warp factor are included and this complicates the effective theory.

Completing the analysis along the lines outlined in this paper could allow for the computation of the finite terms. However, our model in five dimensions with a single scalar cannot capture all features of KKLT. The single extra dimension restricts the fields that can be present in the bulk and a two-form can be sourced in the true theory. Integrating out $\Sigma$, leaving only $\Sigma'$ does not properly address the singular brane potential. Since the four-cycle radius is bigger than the two orthogonal dimensions, we should address any singularities which arise when integrating out $\Sigma$ and leave the four dimensions uncompactified when doing so. 

In principle, we can treat the superpotential for a D7 brane as we did for the 5-dimensional bulk gaugino condensate, namely the superpotential generated by gaugino condensation at low energy. Since the four dimensions are larger than the orthogonal dimensions, we expect the appropriate superpotential at high energy has a large number of ``flavors," corresponding to the KK modes of the gauge bosons and the gauginos. At low energies, we know the fields are integrated out so we expect the low energy condensate scale to have power law $\Sigma'$ dependence, in addition to the exponential. Of course to properly address this and other questions requires a more complete analysis to calculate the finite result if and when singularities are canceled. Our model  as presented highlights some issues but is too simple to resolve them, which we leave to future work.

\section{Supersymmetry Breaking and Anomaly Mediation}

\label{sec:AMSB}

In the preceding sections we outlined the necessary ingredients to generate an AdS$_4$ minimum, which allows for  flat space after including a SUSY-breaking sector on either of the branes. In this section we discuss the masses generated by $\Fcp$ in these scenarios. We also show how to reproduce the 5d result from within a 4d EFT for each of the different models.

In a 4d theory, anomaly mediation can be derived in terms of a single scale, $\Fcf/C_4 \simeq \Lambda_\ads^2/\MP$, which sets the size of SUSY-breaking masses for \textit{all} fields -- up to the $\beta$-functions of the low-energy theory. 
Even if originating from a higher-dimensional theory, this would argue that anomaly mediation should be independent of whether fields originated in the UV or the IR, for example. They would get the anomaly-mediated mass predicted in the 4d theory in terms of the AdS scale, with no warp factor, 5d stabilization parameter, or other potential suppression. 

In this section we will show that this is not the case in general. Anomaly-mediated masses for IR fields are warped, as we will see is also the case for tree-level masses in AdS$_4$ space. Furthermore, in $\beta$-models, the anomaly-mediated masses are suppressed by the size of the no-scale breaking, which is set by $\beta$. It has been argued there there can be  non-universal anomaly-mediated masses originating from a higher-dimensional theory (see, e.g.~\cite{Luty:2000ec, Luty:2002ff, Arkani-Hamed:2004zhs}). In refs.~\cite{Luty:2002ff,Arkani-Hamed:2004zhs} for example, a model was presented in which separate superpotentials on the UV and IR boundaries set independent anomaly-mediated masses for fields on each brane. We found that such boundary potentials can generate such non-universal anomaly-mediated masses, but only when the no-scale structure is broken. Furthermore, unless there is a bulk superpotential, there is no source of negative energy. In all models that led to an AdS$_4$ solution, $\Fcp$ was sourced by a superpotential that was constant throughout the bulk. This bulk superpotential, like a 4d superpotential, generates anomaly-mediated masses for all fields. However, the IR masses are warped, unlike the expectation from the 4d EFT. 

To study how SUSY-breaking is communicated to fields on the boundaries, we include chiral fields $Q_0, Q_\pi$ on each brane. We ignore gauge interactions, although the extension to more general boundary superpotentials is straightforward. We also assume there are no couplings of bulk fields to $Q_\alpha$. The boundary Lagrangians in that case take the form
\begin{align}
    \mathcal{L}_{\rm bdy} &= \sum_{\alpha = 0, \pi} \delta(y-\alpha)
    \left[ e^{2\sigma} \dfour |\C_+|^{4/3} f_\alpha (Q^\dag_{\alpha}, Q_{\alpha})
    +  e^{3\sigma} \left( \dsq \C_+^{2} W_\alpha (Q_{\alpha}) + h.c. \right)\right] \, .
    \label{eq:bdy_theory}
\end{align}

\subsection{Supersymmetry Breaking from $\Fcp$}

Before studying anomaly-mediation, we first note that in our AdS$_4$ solutions the stabilization sector itself necessarily breaks supersymmetry, even without SUSY-breaking on the boundaries. 
This is due to the warp factor dependence of $\Fcp$, which means that scalars and fermions from chiral multiplets on the IR brane do not have the appropriate mass splittings to preserve supersymmetry in AdS$_4$.

We first consider what happens in the condensate model at the AdS$_4$ minimum, where the AdS scale $\Lambda_\ads$ was related to the compensator by (see eq.~\eqref{eq:ads_min})
\begin{align}
    \Lambda^4_\ads &= \frac{\left | \langle e^{- \sigma} \Fcp \rangle \right |^2 }{3k} \, ,
\end{align}
or equivalently
\begin{align}
    \Fcp &= \sqrt{3k} e^{\sigma} \Lambda^2_\ads  \, .
\end{align}
The $\Fcp^2 |q_\alpha|^2$ terms then lead to masses for the scalar components, which are given by
\begin{align}
    &m^2_{q_0} = -\frac{2\Lambda^4_\ads}{3\MP^2} \, ,
    &&m^2_{q_\pi} = -\frac{2\Lambda^4_\ads}{3\MP^2} e^{-2 k \pi r} \, ,
\end{align}
while the fermion remains massless if there is no mass term in the boundary superpotential. In order to preserve supersymmetry in AdS$_4$, the scalar components of a chiral multiplet should get mass terms equal to $m_q^2 = - 2\Lambda^4_\ads/(3\MP^2)$. When combined with an explicit mass term in the superpotential, which is necessary to give positive masses to the $q$'s, this will lead to the SUSY-preserving mass splittings in AdS space (see refs.~\cite{Dine:2007me, Gripaios:2008rg, Rattazzi:2009ux, DEramo:2012vvz, DEramo:2013dzi} for related discussions). 

We therefore see that fields in the UV have the correct mass contributions to preserve supersymmetry, but due to the warp factor dependence the masses of the IR fields break supersymmetry. This means that SUSY is broken in AdS$_4$ space even before adding an explicit SUSY-breaking sector -- if the extra dimension is warped. Stabilizing moduli generally requires moving away from flat space so we expect that most such models break supersymmetry through the back-reaction on the metric.

\subsection{Anomaly Mediation in 5d}

\label{subsec:AMSB5d}
In this section we return to 4d Minkowski space and derive the anomaly-mediated masses from the 5d theory, before presenting the 4d effective theory that matches those results. 

The models in which the no-scale structure was broken by a condensate lead to AdS$_4$ solutions, with a 4d AdS scale $\Lambda_\ads$ and $\Fcp$ given by
\begin{align}
    & \Lambda^4_\ads = \frac{3M_5^5}{4k} \left |W_{\rm bulk} \right |^2 \, ,
    &&\Fcp = \sqrt{3k} \, \Lambda^2_\ads e^{\sigma(y)} \, . 
    \label{eq:FC_ads4_cond}
\end{align}
In $\beta$-models, a bulk superpotential also leads to an AdS$_4$ solution after the no-scale form of the K\"ahler potential is broken by $\beta \neq 0$. In this case $\Fcp $ is suppressed by $\beta$ and inverse powers of $r$:
\begin{align}
    & \Lambda^4_\ads = \frac{M_5 \beta}{kr^4} \left |W_{\rm bulk} \right |^2 \, ,
    &&\Fcp = \sqrt{\frac{4\beta k}{3(M_5 r)^4}} \, \Lambda^2_\ads e^{\sigma(y)} \, .
    \label{eq:FC_ads4_hyper}
\end{align}
In both models we have 
\begin{align}
    \Fcp = c \sqrt{k} \, \Lambda^2_\ads e^{\sigma(y)} \, , 
    \label{eq:c_defn}
\end{align}
where $c =\sqrt{3}$ for the condensate models, and ${c = (M_5 r)^{-2}\sqrt{4\beta/3}}$ in the hypermultiplet models. Below we will use $\Fcp = c \, \Lambda^2_\ads e^{\sigma(y)} $ and treat $c$ as a model-dependent parameter. The anomaly-mediated masses are then proportional to $M^{(AM)}_\alpha$ times the beta-functions of the low-energy theory, where
\begin{align}
    M^{(AM)}_\alpha = \frac{\Fcp(y_\alpha)}{C_+(y_\alpha)} = \frac{c \, \Lambda_\ads^2}{\MP} e^{\sigma(y_\alpha)} 
\end{align}
Taking $f_\alpha=Q_\alpha^\dagger Q_\alpha$ for simplicity, we will be left with a flat 4d theory after SUSY-breaking terms are added if we require that the AdS$_4$ scale satisfies
\begin{align}
    \Lambda_{\rm{AdS}_4}^4 = \frac{k}{M_5} \left( |F_{Q_0}|^2 +e^{-4k\pi r} |F_{Q_\pi}|^2\right) \, .
    \label{eq:FC_scale}
\end{align}
For both models this tuning fixes the scale of the bulk compensator $F$-term.\footnote{Here we assume that $F$- and $D$-terms for bulk fields vanish. If there is SUSY-breaking from bulk fields, there would be additional terms on the right hand side of~\eqref{eq:FC_scale}.}

If we consider the case of a gaugino mass for concreteness, gauginos $\lambda_0$ on the UV brane will have a mass term
\begin{align}
    &\mathcal{L} \sim \int dy \, \delta(y)\frac{\beta(g)}{2g} \frac{\Fcp}{C_+}  \lambda_0\lambda_0 =  m_{\lambda_0} \lambda_0\lambda_0 \, , 
    &&m_{\lambda_0} =  \frac{c \beta(g)}{2g} \frac{\Lambda_\ads^2}{\MP} \, ,
\end{align}
where we have used $\MP = M_5^{3/2}/k^{1/2}$. For $c\sim \mathcal{O}(1)$, as in the condensate models, this matches the expectation from the 4d theory that masses should scale as $\Lambda^2_\ads/\MP$. Bulk gauginos will get masses that similarly match the 4d expectation when $c\sim \mathcal{O}(1)$. The $\beta$-models do not agree with the 4d expectation, however, as $c \ll 1$ in these models. As SUSY is already broken in 5d, so there must be additional (non-supersymmetric) matching terms that need to be included in the 4d theory in order to reproduce the 5d result. We discuss this in more detail in the next subsection.

For gauginos $\lambda_\pi$ on the IR brane the anomaly-mediated masses are:
\begin{align}
    &\mathcal{L} \sim \int dy \, \delta(y - \pi)\frac{\beta(g)}{2g} \frac{\Fcp}{C_+}  \lambda_\pi \lambda_\pi =  m_{\lambda_\pi} \lambda_\pi \lambda_\pi  \, , 
    &&m_{\lambda_\pi} =  \frac{c \beta(g)}{2g} \frac{\Lambda^2_\ads}{\MP}  e^{-k\pi r_c}\, .
\end{align}
The masses for IR fields are warp-factor suppressed, just like any other mass term, due to the warp-factor dependence of $\Fcp$. The masses for IR fields therefore disagree with the 4d expectation due to the extra warp factor dependence, as was the case for the tree-level masses in AdS$_4$ space that we discussed in the previous section.

\subsection{Anomaly Mediation in 4d EFT}

In this section we demonstrate how to construct 4d effective theories that 
reproduce the results of the 5d analysis. As before, we are not directly deriving the 4d theories 
but finding the appropriate superpotentials to match the 5d results. Crucial to our results in the previous sections was the breaking of the no-scale structure. Either a superpotential from a gaugino condensate breaks no-scale or loop corrections to the K\"ahler potential do so. As this breaking occurs in very different ways in both models, we  separately consider the effective theories that reproduces the anomaly-mediated masses derived in 5d, which look very different in the two cases.

\subsubsection{Condensate Models}

We consider condensate models where the constant superpotential is only in the bulk, and take the limit $\alpha M_5 r \gg1$. This limit simplifies the expressions and amounts to assuming that the extra dimension is stabilized at large $r$. The 4d effective superpotential, $W_{\rm eff, \, gc }$, is then given by (see eq.~\eqref{eq:Weff_gc}):
\begin{align}
    W_{\rm eff, \, gc } (\rho)&= 
   \sqrt{\frac{k}{M_5}} \left[ \frac{W_{\rm bulk}}{4}(1 + 3 \rho )
    + \frac{\alpha W_\lambda }{1+x} \rho^{1+x} \right]\, .
\end{align}
The leading (constant) term in $\Fcf$ is then
\begin{align}
    \Fcf&= \frac{M_5^{5/2} W_{\rm bulk}}{4 \sqrt{k}} 
    = \frac{\Lambda^2_\ads  }{2\sqrt{3}} \, .  
    \label{eq:FC_gc_EFT}
\end{align}
It only agrees with the 5d result~\eqref{eq:gc_Fs} at the minimum of the potential, where $3\alpha W_\lambda \rho^x = -4 W_{\rm bulk}$, but this is sufficient to get the right anomaly-mediated masses for elementary fields on the UV brane or in the bulk. This $F_{C_4}$ does lead to unwarped masses for IR localized fields, however, so fails to reproduce the 5d result in this respect. 

This can be rectified by adding a $\rho^3$ term to $W_{\rm eff, \, gc }$, so that the effective superpotential now reads:
\begin{align}
    \widetilde{W}_{\rm eff, \, gc } (\rho)&= 
   \sqrt{\frac{k}{M_5}} \left[ \frac{W_{\rm bulk}}{4}(1 + 3 \rho + 6 \rho^3)
    + \frac{\alpha W_\lambda }{1+x} \rho^{1+x} \right]\, ,
    \label{eq:Wgc_AM}
\end{align}
This term is negligible when determining the stabilizing potential, being suppressed by extra powers of $\rho$ relative to the other terms. It will, however, allow us to derive the warped masses for IR-localized fields.
To show this, we include chiral fields on the UV and IR branes, so that the full 4d EFT describing this theory is:
\begin{align}
    \frac{\mathcal{L}}{\sqrt{-g}} &= \dfour |\C_4|^2 \left( -3 + 3 \left | \rho \right |^2 + \tilde f_0 \left( Q_0^\dag,  Q_0 \right) + \left | \rho \right |^2 \tilde f_\pi \left( Q_\pi^\dag,  Q_\pi \right) \right) 
    + \left[ \dsq \C_4^3 \widetilde{W}_{\rm eff, \, gc } (\rho) 
    +h.c. \right] \, .
\end{align}
The functions $\tilde f_{0, \pi}$ are given by rescaling the functions in the 5d theory (see eq.~\eqref{eq:bdy_theory}) by a factor of $M_5/k$ to account for the different scalar component of $\C_4$ compared to $\C_+$. If we then make the field redefinition $\omega = \rho \, \C_4$, as proposed in refs.~\cite{Luty:2000ec, Luty:2002ff}, we find
\begin{align}
    \frac{\mathcal{L}}{\sqrt{-g}} &= \dfour \left[ |\C_4|^2 \left( -3 + \tilde f_0 \left( Q_0^\dag,  Q_0 \right) \right) + |\omega|^2 \left(3 +  \tilde f_\pi \left( Q_\pi^\dag,  Q_\pi \right)\right) \right] + \left[ \dsq  \C_4^3 \, \widetilde{W}_{\rm eff, \, gc } \left(\omega/\C_4 \right) +h.c. \right] \, . 
     \label{eq:gc_EFT}
\end{align}

In ref.~\cite{Luty:2002ff} it was argued that fields on the UV brane have anomaly-mediated masses set by $\Fcf/C_4$, while fields on the IR brane have anomaly-mediated masses set by $F_\omega/\omega$. With the superpotential~\eqref{eq:Wgc_AM}, the IR brane masses are then set by 
\begin{align}
    \frac{F_\omega}{\omega} = \frac{F_\rho}{\rho} + \frac{\Fcf}{C_4} \simeq -\frac{3\rho}{2} M_5 W_{\rm bulk} = \frac{\sqrt{3}\rho}{\MP}  \Lambda^2_\ads  \, ,
\end{align}
thus reproducing the 5d result of warped anomaly-mediated masses also for fields on the IR brane. This EFT also captures the fact that SUSY is broken, as $F_\rho$ is a SUSY-breaking order parameter and is necessarily nonzero (as was also noted in ref~\cite{DEramo:2012vvz}) and is in fact comparable in magnitude to $\Fcf/C_4$. 

Refs.~\cite{Luty:2000ec, Luty:2002ff} proposed the $\C_4^3 - \omega^3$ form of the superpotential as arising due to separate superpotentials on each of the branes. Here we see that this form is needed to reproduce the warped anomaly-mediated masses of the 5d theory. In the analysis above these mass terms are sourced by a superpotential in the bulk. The effective theory described by equation~\eqref{eq:gc_EFT} fully captures the stabilized 5d theory, giving the right stabilizing potential and  cosmological constant.  This also gives the correct anomaly-mediated masses generated by a bulk superpotential for fields on both branes. The full set of anomaly-mediated mass terms will include both bulk and boundary contributions, as we show in the next section.

\subsubsection{Including Boundary Superpotentials}

Our derivation of the effective theory has so far ignored possible boundary superpotential terms of the type referred to above. It is relatively straightforward to include a superpotential on the UV brane by adding the term
\begin{align}
    W_{\uv} (\rho, Q_0)&= \left( \frac{k}{M_5} \right)^{3/2} W_0 (Q_0) (1 + 3 \rho + 6 \rho^3) \, ,
    \label{eq:WUV}
\end{align}
which is the same term as in equation~\eqref{eq:Weff_gc} except we now allow for field-dependence in $W_0$ and add a $\rho^3$ term so that $F_\omega$ has the correct warp factor dependence. With the inclusion of $W_\uv$, the anomaly-mediated mass scales for UV and IR fields are given by (up to exponentially small corrections)
\begin{align}
    \frac{F_\omega}{\omega} &= -\frac{3\rho}{2} \left(4k \langle W_0 \rangle  + M_5 W_{\rm bulk} \right) = \frac{\sqrt{3}\rho}{\MP}  \Lambda^2_\ads  \, , \nonumber
    \\
    \frac{\Fcf}{\MP} &= k \langle W_0 \rangle  + \frac{M_5}{4} W_{\rm bulk} = \frac{\Lambda^2_\ads  }{2\MP\sqrt{3}}
    \, .
\end{align}
Again these mass scales match the 5d theory at the minimum of the potential, as $3\alpha W_\lambda \rho^x = -4 \left(W_{\rm bulk} + 4k\langle W_0 \rangle/M_5 \right)$. For this reason it makes sense that IR-localized fields have masses proportional to $W_0$, despite the fact that any $W_0$ term in the 5d theory will come with a $\delta(y)$ factor and decouple from fields in the IR.

Including a superpotential on the IR brane is slightly more involved, as in 5d $W_\pi$ does not contribute to the potential so should not appear in the anomaly-mediated masses. However, including the term
\begin{align}
    \rho^3 W_{\ir} (Q_\pi)&= \rho^3 \left( \frac{k}{M_5} \right)^{3/2} W_\pi (Q_\pi) \, ,
\end{align}
which gives the correct potential for fields on the IR brane, also contributes to $F_\omega$ at the level
\begin{align}
    \frac{F_\omega}{\omega} &\sim - M_5 \langle W_\pi \rangle \rho \, .
\end{align}
This leads to additional mass terms for IR fields which are not present in the 5d theory. The discrepancy comes about as in the 5d theory $\Fcp$ is set by $W_\lambda \rho^x$, which is unrelated to $W_\pi$ because the $W_\pi$ terms are exponentially small corrections to the potential. To generate only the interaction terms involving the IR fields, without leading to unwanted contributions to $F_\omega$, we can redefine $W_\ir$ so it has vanishing expectation value:
\begin{align}
    W_{\ir} (Q_\pi)&= \left( \frac{k}{M_5} \right)^{3/2} \left( W_\pi (Q_\pi) - \langle W_\pi \rangle \right) \, .
    \label{eq:WIR}
\end{align}

The full EFT, including a bulk superpotential and superpotentials on both branes, is then given by
\begin{align}\label{eq:gc_EFT_full}
    \frac{\mathcal{L}}{\sqrt{-g}} &= \dfour \left[ |\C_4|^2 \left( -3 + \tilde f_0 \left( Q_0^\dag,  Q_0 \right) \right) + |\omega|^2 \left(3 +  \tilde f_\pi \left( Q_\pi^\dag,  Q_\pi \right)\right) \right] 
    \\
    &\hspace{5mm}+ \left[ \dsq  \C_4^3 \, \left( \widetilde{W}_{\rm eff, \, gc } \left(\omega/\C_4 \right) + W_\uv(\omega/\C_4, Q_0) \right) + \omega^3 W_{\ir} (Q_\pi) + h.c. \right] \, , \nonumber
\end{align}
with $W_\uv$, $W_\ir$ and $\widetilde{W}_{\rm eff, \, gc }$ defined in equations~\eqref{eq:WUV}, \eqref{eq:WIR} and~\eqref{eq:Wgc_AM} respectively. In the 5d theory $F_\Sigma$ had terms proportional to $\delta (y) W_0$ and $\delta (y-\pi) W_\pi$, and any SUSY-breaking masses generated from these terms need to be added to the EFT as explicit SUSY-breaking terms.

\subsubsection{$\beta$-Models}

Constructing an effective theory which captures the effects of anomaly mediation is more difficult for the models considered in sections~\ref{sec:hypermultiplets} and~\ref{sec:beta_models}, where the extra dimension was stabilized by hypermultiplets and the no-scale structure broken by loop corrections. The reason for this is that $\Fcp$ is suppressed relative to the AdS$_4$ scale, or in the parametrization of eq.~\eqref{eq:c_defn} that $c\ll 1$. This is because supersymmetry is broken by $F_\Sigma$ and the dominant contribution to the AdS energy is from the cross term $F_\Sigma \Fcp$.

This means that a supersymmetric EFT will not capture the full details of the 5d model. In particular, for warped extra dimensions it will not reproduce the effective potential and anomaly-mediated masses at the same time. The effective theories constructed in sections~\ref{sec:hypermultiplets} and~\ref{sec:beta_models} were designed to reproduce the stabilizing potential of the 5d theory, but fail to reproduce the correct anomaly-mediated masses. This can be seen as the effective superpotential~\eqref{eq:Weff3} leads to $\Fcf \propto \sqrt{\beta} W_{\rm bulk}$, which does not match the $\beta$ dependence of the 5d theory, where $\Fcf \propto \beta W_{\rm bulk}$. In this section we construct a different effective theory, that reproduces the anomaly mediation of the 5d theory. This comes at the cost of not giving the right stabilizing potential or the AdS$_4$ cosmological constant, which will need to be added to the theory as a extra, non-supersymmetric terms.

The full Lagrangian we consider is
\begin{align}
    \mathcal{L}_{4d} = \mathcal{L}_{0} (Q_0) + \mathcal{L}_{\pi} (Q_\pi, \rho) -  V_{\rm match} (Q_0, Q_\pi) - V_{\rm stab}(\rho) - V_0 \, ,
    \label{eq:AM_EFT}
\end{align}
where:
\begin{itemize}
    \item $\mathcal{L}_{0}$ ($\mathcal{L}_{\pi}$) are Lagrangians for fields on the UV (IR) branes, which we will construct in a supersymmetric formalism so that they reproduce the anomaly-mediated masses of the 5d theory;
    \item $V_{\rm match}$ includes SUSY-breaking terms that need to be included due to a nonzero $F_\Sigma$ or from boundary superpotentials;
    \item $V_{\rm stab}$ is the stabilizing potential, given by eq.~\eqref{eq:Veff_J2}; and
    \item $V_0$ is constant term introduced so that $\mathcal{L}_{4d}$ gives the correct cosmological constant. $V_0$ contains a term which gives the AdS$_4$ energy density in eq.~\eqref{eq:DeltaV_beta}, plus $\mathcal{O}(\beta^2)$ corrections to cancel any constant potential terms coming from $\mathcal{L}_{0} + \mathcal{L}_{\pi}$.
\end{itemize}
Dividing the theory up in this way is most relevant when anomaly mediation is the main communicator of SUSY-breaking, i.e. in models where the terms in $V_{\rm match}$ are smaller than the anomaly-mediated contributions. Then the dominant SUSY-breaking terms come from anomaly mediation and can be captured by $\mathcal{L}_0$ and $\mathcal{L}_\pi$, which are given by:
\begin{align}
    \frac{\mathcal{L}_{0} (Q_0)}{\sqrt{-g}} &= \dfour  |\C_4|^2 \left( -3 + Q_0^\dag Q_0 \right) + \left[ \dsq  \C_4^3 \left( \sqrt{\frac{4k}{3}} \frac{\beta W_{\rm bulk}}{M_5^{5/2}r^4 } + W_0(Q_0) \right) +h.c. \right]\, , \nonumber
    \\
    \frac{\mathcal{L}_{\pi} (Q_\pi, \rho)}{\sqrt{-g}} &= \dfour  |\omega|^2 \left(3 +  Q_\pi^\dag Q_\pi \right) 
    + \left[ \dsq   \omega^3 \left( \sqrt{\frac{4k}{3}} \frac{\beta W_{\rm bulk}}{M_5^{5/2}r^4 } + W_\pi(Q_\pi)  \right)+h.c. \right] \, , 
\end{align}
where we have again made the field redefinition $\omega = \C_4 \rho$ and the constant term in both superpotentials is chosen to reproduce the 5d anomaly-mediated masses for the hypermultiplet model.

As in the EFT constructed for the condensate models, $\Fcf/C_4$ sets the anomaly-mediated masses in the UV, while $F_\omega/\omega$ sets them in the IR. $F_\omega/\omega$ is proportional to $\omega$, reproducing the warped anomaly-mediated masses found in the 5d theory. SUSY-breaking is indicated both by the fact that $F_\rho \neq 0$, as well as the explicit SUSY-breaking terms in $V_{\rm match}$ and $V_{\rm stab}$. The dominant source of SUSY-breaking in this type of models is model-dependent, but for models where the terms in $V_{\rm match}$ are small or can be treated independently of the anomaly-mediated contributions the parametrization of eq.~\eqref{eq:AM_EFT} is a useful way to capture the dominant SUSY-breaking effects.

\section{Summary}

\label{sec:EFT_summary}

Because of the many subtleties and cases that have led to distinct 4d effective theories, many of which require the addition of non-supersymmetric matching terms to reproduce the 5d theory, we summarize our results here. To consolidate the results, we show the dependence of $\Fcp$ and $F_\Sigma$ on the constant (or radion-dependent) superpotentials in Table~\ref{tab:F_summary}, treating separately the different types of no-scale breaking. Note we are neglecting any possible contributions from an explicit supersymmetry breaking sector. 

An important ingredient has been how the no-scale form of the bulk K\"ahler potential is broken. When preserved, the no-scale structure gives a small or zero compensator $F$-term, leading to a potential that would be minimized at $V=0$ when SUSY is unbroken with $\Fcp = 0$. This would lead to de-Sitter space when SUSY-breaking sectors on the boundaries are included. This means that the no-scale structure should be broken in order to generate an AdS$_4$ minimum in the absence of SUSY breaking terms, so that we ultimately end in flat space. A simple stabilization mechanism with boundary sources generates a nonzero potential for the radion, but doesn't change the equation of motion for the radion $F$-term ($F_\Sigma$), leaving the no-scale structure essentially intact. Such models can lead to no-scale supersymmetry-breaking through $F_\Sigma$  in the presence of bulk or boundary constant superpotentials, but would not lead to anomaly mediation unless the no-scale structure is broken. 

One way this breaking can occur is through loop corrections to the K\"ahler potential and a constant bulk superpotential, $W_{\rm bulk}$. In this case, illustrated in the first row of Table~\ref{tab:F_summary}, both $\Fcp$ and the AdS$_4$ scale are non-zero but suppressed by the size of the loop corrections, whereas $F_\Sigma$ is nonzero and unsuppressed by loop corrections. Such a model generates anomaly-mediated masses from $\Fcp$ for all fields, regardless of where in the extra dimension they are localized, due to $\Fcp$ being sourced by a constant superpotential in the bulk. However, these masses are proportional to the warp factor and can be subleading in the IR relative to other loop corrections~\cite{Luty:2002ff, Arkani-Hamed:2004zhs}, which is not apparent from the na\"ive 4d EFT.

As the next two rows of Table~\ref{tab:F_summary} illustrate, $F_\Sigma$ and $\Fcp$ (which is still $\beta$-suppressed) can have contributions from  boundary superpotentials too, but these contributions are restricted to the boundary where they are sourced.  
In the presence of boundary superpotentials, there are therefore non-universal anomaly-mediated masses, which contribute to masses only for fields with support on the same brane where $W$ is nonzero. Furthermore, boundary superpotentials modify the boundary conditions satisfied by bulk fields in such a way that the potential would still be minimized for $V=0$ in the absence of a bulk superpotential. 

A supersymmetric 4d EFT does not naturally reproduce these effects, since a nonzero $\Fcf$ would generate both universal anomaly-mediated masses and negative energy density, which is not true for the 5d theory. The boundary potentials act as sources for the bulk fields and also a localized $\Fcp$, but don't lead to negative energy density. Furthermore, the supersymmetric EFT in these models is not able to reproduce both the stabilizing potential and the anomaly-mediated masses. We showed how to construct a supersymmetric EFT that captures either one of these features, with the other terms needing to be added as matching terms.
Because the nonzero $F_\Sigma$ breaks supersymmetry in the 5d theory for these models, it is simplest to include explicit SUSY-breaking terms in the 4d theory, as is is not obvious how to capture such masses in an EFT in which SUSY is solely broken spontaneously.

An alternative way of breaking the no-scale form is with a $\Sigma$-dependent superpotential in the bulk, $W(\Sigma)$ (for example from gaugino condensation). Models stabilized by condensates lead to an AdS$_4$ minimum when a constant ($\Sigma$-independent) superpotential is also included. If the constant superpotential term is added to the bulk then $F_\Sigma =0$, while if it is added to the branes $F_\Sigma$ is nonzero locally, but $\int dy \, F_\Sigma =0$, both of which are illustrated in the final row of Table~\ref{tab:F_summary}. This means that for a nonzero boundary superpotential, only fields with non-trivial bulk profiles that couple to $F_\Sigma$ will get SUSY-breaking masses, so a gaugino with a flat profile would remain massless at leading order. In these models $\Fcp$ is related to $W'(\Sigma)$, which is constant throughout the bulk, so  generates anomaly-mediated masses for all fields regardless of where they are localized. This agrees with the expectation from the 4d theory, although unlike the simple 4d theory these masses are warped. 
We showed in section~\ref{sec:AMSB} how to construct a supersymmetric 4d EFT that matches the 5d results, which was not possible in the $\beta$-models. In such models for which matching terms are absent or subleading we expect results to qualitatively match those of conformally sequestered models~\cite{Luty:2001jh,Luty:2001zv,Sundrum:2004un, Schmaltz:2006qs}.

Many of the features highlighted above do not emerge from a supersymmetric 4d analysis. As an example, the SUSY-breaking terms from $F_\Sigma$ and the boundary terms in $\Fcp$ must be added as explicit supersymmetry-breaking matching contributions as they  do not arise from a 4d supersymmetric theory. The source of the inadequacy of a purely 4d analysis is the supersymmetry breaking in five dimensions. We therefore review the sources of supersymmetry breaking we have found.

In models with AdS$_4$ vacua, supersymmetry-breaking is manifest in the warped spectrum generated by $\Fcp$, which generates IR masses that don't have the mass splittings required to preserve supersymmetry in AdS$_4$. In 4d Minkowki space (after adding explicit supersymmetry breaking to get zero energy), the anomaly-mediated mass terms are still warped. We showed how to capture this in a supersymmetric EFT, where the SUSY-breaking is manifested in the nonzero $F_\rho$. Though not relevant to our 4d EFTs that we construct in flat space, additional sources of SUSY-breaking in AdS$_4$ are  the nonzero $D$ terms for the graviphoton and $\Fcm$, which are proportional to the AdS$_4$ scale.

A further source of supersymmetry breaking is that $F_\Sigma$ is nonzero in $\beta$-type models with a constant superpotential, $W_{\rm bulk}$, meaning supersymmetry is broken at a scale set by $W_{\rm bulk}$. On top of this, boundary terms in these models also break supersymmetry and generate nonzero $F_\Sigma$ and $\Fcp$.
Although bulk superpotentials do not induce $F_\Sigma$ in condensate models, boundary superpotentials do. So with boundary superpotentials additional SUSY-breaking terms would need to be included in condensate models as well. We find there is no simple way to construct a fully supersymmetric 4d EFT that captures these features of the 5d theory, and these additional supersymmetry-breaking masses need to be incorporated as matching contributions.  

Theories that break supersymmetry only in the low-energy EFT and are consistent with a supersymmetric 4d description are the exception. Examples include supersymmetry-breaking in the IR, which leads to an unwarped anomaly-mediated mass spectrum in the UV (albeit with warped anomaly-mediated masses in the IR), and models with a constant bulk superpotential and no-scale breaking through a $\Sigma$-dependent bulk superpotential.

Aside from supersymmetry-breaking, the other obstacle to a simple 4d effective theory is the 4d radion potential associated with a 5d warped theory, which does not reflect the  no-scale structure. The superpotential, $W_{\rm eff}$, in the 4d effective theory can therefore not simply be the integral of the  bulk superpotential over the extra dimension (as has been previously assumed).

\begin{table}[h!]
\centering

\renewcommand{\arraystretch}{1.5}
\begin{tabular}{|c|c|c|c|c|}
\hline
\begin{tabular}[c]{cc}
 No-scale   \\ Breaking  
\end{tabular}    & 
\begin{tabular}[c]{cc}
 Constant   \\ Superpotential  
\end{tabular} & $F_{\Sigma}$ & $ \Fcp$   &   $\Lambda^4_\ads$\\ 
\hline
\hline
    K\"ahler & Bulk & $ - M_5^2 r W_{\rm bulk}\, e^{\sigma (y)} $ & $ -  \frac{ 2 \beta}{M_5^{3/2} r^4}  W_{\rm bulk} e^{\sigma (y)}$   &    \multirow{3}{*}{$\frac{M_5 \beta}{kr^4} \left |W_{\rm bulk} \right |^2$}\\ 
\cline{2-4}
\begin{tabular}[c]{@{}l@{}}
 Potential  
\end{tabular} 
    & UV      &  $- 2 M_5 W_0  \delta(y)$  & $ - \frac{18 \beta }{M_5^{3/2} r^4}  W_0 \delta(y) $   &  \\ 
\cline{2-4}
     & IR      &  $- 2 M_5 W_\pi e^{\sigma(\pi)}\delta(y-\pi)  $   & $ - \frac{18 \beta e^{\sigma(\pi)}}{M_5^{3/2} r^4}   W_\pi\delta(y-\pi)$  &  \\ 
\hline
Superpotential &  Bulk $+$ UV   &    $- M_5 W_0  \left( 2\delta(y) - k r e^{- k r y}  \right)$
& $\alpha W_{\lambda} e^{\sigma (y) - M_5 r\alpha} M_5^{5/2} $   & $\frac{3M_5^5}{4k} \left |W_{\rm bulk} \right |^2$\\
\hline
\end{tabular}
\caption{Summary of the leading expressions for $F_\Sigma $ and $\Fcp$ for each way of breaking the no-scale structure, with superpotentials in the bulk and/or on the branes, and the AdS$_4$ scale (before adding boundary SUSY-breaking sectors) for each type of no-scale breaking. The relevant no-scale breaking terms are $\Delta \mathcal{L} = \frac{1}{3} \beta e^{2\sigma} \dfour \mathcal{V}_\Sigma^{-3} |\C_+|^{4/3}$ for the K\"ahler models, and the $\Sigma$ dependence in~${W = W_0 \delta(y) + M_5 r W_{\rm bulk} + W_\lambda e^{-2\alpha \Sigma} }$ for the superpotential models. The results when no-scale is unbroken are the $\beta \to 0$  expressions for K\"ahler potential breaking models.
}
\label{tab:F_summary}
\end{table}
As can be seen from Table~\ref{tab:F_summary}, in most models $F_\Sigma$ breaks supersymmetry, with the exception being superpotential no-scale breaking with a superpotentials only in the bulk (so $W_0 = 0$). 
We note that for the case of superpotential no-scale breaking, $F_\Sigma = 0$ exactly when $W_0 = 0$ but $W_{\rm bulk}$ is non-zero, while in general $\int dy F_\Sigma = 0$ even though $F_\Sigma $ is nonzero locally due to $W_0$. 
Bulk fields therefore can have masses both  at tree-level from  $F_\Sigma$ and loop-level contributions from $\Fcp$, whereas $F_\Sigma$ does not couple at tree level to boundary-localized fields. 
As an example of the masses generated we give the formula for gaugino masses, neglecting potential model-dependent loop contributions from $F_\Sigma$. We consider both bulk gauginos, $\lambda_{\rm bulk}$, with a zero-mode wavefunction $\lambda_0(y)$, and boundary gauginos $\lambda_\alpha$, where $\alpha = 0,\pi$:
\begin{align} 
    m_{\lambda_{\rm bulk}} &= \int dy |\lambda_0(y)|^2 \left( F_\Sigma(y)+\frac{\beta(g)}{2g} \frac{\Fcp(y)}{M_5^{3/2}} \right) \, ,
    \\
    m_{\lambda_\alpha} &= \frac{\beta(g)}{2g}\frac{\Fcp(\alpha)}{M_5^{3/2}} \, .
\end{align}

In models where no-scale is broken in the K\"ahler potential, $\Fcp$ is given by the sum of each of the contributions given in Table~\ref{tab:F_summary}, i.e.
\begin{align}
    \Fcp = -  \frac{ 2 \beta  e^{\sigma (y)}}{M_5^{3/2} r^4}  \left[ W_{\rm bulk} + 9 \left( W_0 \delta(y) +  W_\pi\delta(y-\pi) \right) \right] \, ,
\end{align}
while in the condensate models $\Fcp$ is always given by the expression in the final row of Table~\ref{tab:F_summary}. Gauginos localized in the UV or with constant $\lambda_0$ will have masses that have no warp factor dependence, whereas gauginos localized in the IR will have masses that are warp-factor suppressed. These results contrast with the universal prediction for anomaly-mediated masses from the 4d theory, $m_\lambda = \beta(g) \Fcf/(2g \MP)$, which does not capture the warp factor dependence of $\Fcp$, the separate contributions from bulk and boundary superpotentials, or possible additional contributions from a nonzero $F_\Sigma$.

\section{Conclusions}

\label{sec:conclusions}

Extra dimensions are generic in phenomenological string models and are a natural setting for anomaly-mediated SUSY breaking. In this work we performed a complete study of how the 4d EFT emerges from a supersymmetric 5d theory and explored anomaly mediation, revealing many surprising features that were not obvious from a 4d analysis. We used the formalism for supergravity on a 5d orbifold bounded by branes, where the $\mathcal{N}=2$ supersymmetry of the bulk is broken to $\mathcal{N}=1$ by orbifolding. We allowed for warping of the extra dimension to study supersymmetry-breaking and anomaly mediation in 5d, and the corresponding 4d effective theory.

We now understand how two very different statements made about anomaly mediation can both be true. On the one hand, anomaly mediation in the 4d theory was shown to be universal, independent of any high energy details. On the other hand, it has been argued that anomaly mediation in the IR and UV can in principle be independent if they are sourced by independent superpotentials on each brane (although the EFT with independent boundary superpotentials used in the past was derived incorrectly).

We now see how the 5d theory can incorporate both these results. All fields get an anomaly-mediated mass -- albeit one that is proportional to a warp factor -- from a bulk superpotential. The bulk superpotential also sources a negative energy density that cancels the positive energy from SUSY breaking and which, when integrated over the extra dimension, directly corresponds to the superpotential in the 4d EFT. This leads to  anomaly mediation that can be captured in a 4d EFT with a radion. 
On the other hand, independent anomaly-mediated masses are possible in models where $\Fcp$ has brane-localized terms. Such masses are derived in the 5d theory and need to be included in the effective theory as SUSY-breaking matching terms.

These results have important phenomenological implications.  Most notably, because of the warping of masses generated by both $\Fcp$ and $F_\Sigma$, the IR theory can have an approximate globally supersymmetric spectrum even when supersymmetry is broken at a high scale, as suggested in ref.~\cite{Luty:2002ff}. Furthermore, in theories with boundary superpotentials and loop corrections to the bulk K\"ahler term, gaugino mediation generically accompanies anomaly mediation, which can solve the negative slepton mass squared problem for sleptons on a brane coupled to bulk gauge bosons. Furthermore, not all masses from $\Fcp$ and $F_\Sigma$ depend on a single scale, as boundary superpotentials can give contributions of a different overall magnitude.

This means there are several features that can give hierarchical masses even within the context of anomaly mediation. Whenever loop corrections are responsible for breaking no-scale, anomaly-mediated masses are suppressed by the size of the loop corrections, relative to the gravitino mass. Even if there is only a bulk superpotential, warp factors suppress anomaly-mediated masses on the IR brane, while there is no such suppression for UV localized field. In this case, depending on the parameters, other loop corrections may dominate in the IR over anomaly-mediated masses. Third, when there are independent brane superpotentials, there are independent contributions to supersymmetry-breaking masses for fields on each brane. 

Explicitly,
\begin{enumerate}
    \item In $\beta$-models with bulk superpotentials, bulk fields get tree-level supersymmetry-breaking terms through coupling to $F_\Sigma$ and from anomaly mediation. Brane fields get masses from anomaly mediation (or potentially from a non-decoupled supersymmetry-breaking sector).
    \item In $\beta$-models with boundary superpotentials, non-universal anomaly-mediated mass terms are generated for boundary fields.
    \item  In condensate models with a bulk superpotential, $F_\Sigma = 0$, and anomaly mediation communicates SUSY-breaking. 
    \item In condensate models with boundary superpotentials, fields with non-trivial bulk profiles can get SUSY-breaking masses from $F_\Sigma$. Boundary fields get both anomaly-mediated and $F_\Sigma$-dependent masses.
\end{enumerate}
Clearly there is a wide range of hierarchies in supersymmetry-breaking masses that are possible. For example, a Higgs localized to the IR brane might have a mass that is suppressed both by loop factors~($\beta$) and a warp factor. Scalar fields localized in the UV could have very heavy masses. Gaugino masses can depend on whether gauge bosons are in the bulk or on the boundary and which type of no-scale breaking is present. This gives rise to a wide range of phenomenological possibilities that are yet to be explored. These can potentially help address the little hierarchy problem and the problem of negative slepton masses squared, which we leave to future work.

Our results may also have implications for studies of moduli stabilization and SUSY-breaking in string theory. Many such studies take a supersymmetric 4d theory as a starting point. We have shown that there are many ways that supersymmetry may be broken already at the level of the higher dimensional theory in a way that is not obvious from a supersymmetric 4d EFT. It is an interesting question whether these results could shed light on how SUSY may be broken in string models and if they can be made to contribute to  a positive cosmological constant, tasks which we also leave to future work. We considered a toy model for the KKLT scenario, where $W(\Sigma)$ is localized on one of the branes rather than in the bulk, and found singular terms in the potential. Extending our setup to a more complete version of a KKLT-like scenario could be another interesting future direction to connect the high and low energy theories and to see how to calculate finite corrections when the  singular terms are resolved.

\section*{Acknowledgments}

We thank Huy Tran for collaborating on the early stages of this project. We thank Rashmish Mishra and Raman Sundrum for discussions and comments on a draft of the manuscript. We also thank Raphael Flauger, Tony Gherghetta, Andreas Karch, Michele Papucci, Alex Pomarol, Riccardo Rattazzi, Michele Redi, Matthew Reece and Thomas van Riet for useful discussions. We would like to acknowledge GRASP Initiative funding provided by Harvard University. MN is supported by NSF Award PHY-2310717.

\appendix

\section{Gauge Fixing Conditions}

\label{app:fermiongauge}

In this appendix we quote the gauge-fixing conditions we impose, referring the reader to refs.~\cite{Kugo:2002js, Abe:2004ar} for more details. As discussed in section~\ref{subsec:gauge}, two conditions fix the scalar components of the compensator fields:
\begin{align}
    &C_- = 0 \, ,
    &&C_+= \sqrt{M_5^3 + |\varphi_+|^2 +|\varphi_-|^2 } \, .
\end{align}
These break the $SU(2)$ symmetry and scale invariance of the bulk theory. The scalar component of the radion was also fixed to be $\varphi_\Sigma = M_5 r$ to get the correct coefficient for the Einstein-Hilbert term. Conformal supersymmetry is broken by setting:
\begin{align}
    &\Omega^2_{R} = 0 \, .
\end{align}
It also relates the fermionic component of the compensator $\chi_{C_\pm}$ to the fermionic components of the hypermultiplets ($\chi_\pm$) by:
\begin{align}
    &C_+\chi_{C_+} = \varphi_ - \chi_{+} +  \varphi_+ \chi_{+} \, .
\end{align}
Special conformal transformations are broken by setting the auxiliary field from the Weyl multiplet to zero:
\begin{align}
    b_M = 0 \, .
\end{align}
The other auxiliary fields of the gravitational multiplet -- $v_{\mu \nu}, v_{\mu y} , B^{1, 2}_\mu, B^3_\mu$ -- are not fixed by any gauge choice, but must all vanish in order to preserve 4d Poincare invariance.

If there are additional matter fields in the bulk then these conditions may be modified. We refer the reader to section 3 of ref.~\cite{Abe:2004ar} for the general expressions. 

\section{Simplifying the Bulk Lagrangian}

\label{app:IBP}

In this appendix we provide more details on determining the boundary terms derived in section~\ref{subsubsec:hypL}. These terms come from integrating the bulk kinetic terms (eq.~\eqref{eq:Lkin}) by parts to put the kinetic terms into a canonical form. Keeping only those terms quadratic in the fields the terms from $\mathcal{L}_{\rm F}$ are:
\begin{align}
    \mathcal{L}_{\varphi_\pm}  =& - \frac{2 e^{4\sigma}}{r}  \left | \left( \partial_y + \frac{3\dot \sigma}{2} \pm g_h M_5 r \right) \varphi_\pm \right |^2 \, .
\end{align}
Expanding and integrating by parts gives
\begin{align}
    \mathcal{L}_{\varphi_\pm} =& -\frac{2e^{4\sigma}}{r}\left[ |\dot \varphi_\pm|^2 +| \varphi_\pm|^2 \left(\frac{3\dot \sigma}{2}\pm rM_5 g_h\right)^2
    + \partial_y |\varphi_\pm|^2 \left(\frac{3\dot \sigma}{2} \pm rM_5 g_h\right)\right ] \, ,
    \\
    =& -\frac{2e^{4\sigma}}{r}\left[ |\dot \varphi_\pm|^2 +| \varphi_\pm|^2 \left(rM_5 g_h \right)^2
    - |\varphi_\pm|^2 \left(\frac{3\ddot \sigma}{2} + \frac{15\dot \sigma^2}{4} \pm  rM_5 (\dot g_h + g_h\dot \sigma) \right)\right ] \, ,
     \\
    =& -\frac{2e^{4\sigma}}{r}\left[ |\dot \varphi_\pm|^2 +| \varphi_\pm|^2 \left(rM_5 g_h \right)^2
    \mp  rM_5 (\dot g_h + g_h\dot \sigma)  |\varphi_\pm|^2 \right ] + \frac{3re^{4\sigma}}{8} \mathcal{R} |\varphi_\pm|^2 \, ,
\end{align}
where we have identified the Ricci scalar as
\begin{align}
    \mathcal{R} = \frac{4}{r^2} \left( 2 \ddot \sigma + 5 \dot \sigma^2 \right) \, .
    \label{eq:Ricci}
\end{align}
The compensator Lagrangian has the same form as the hypermultiplet Lagrangian with an overall minus sign. The analogous terms for the compensator, $C_+$, can thus be obtained by taking $\mathcal{L}_{\varphi_+}$, multiplying by an overall minus sign and substituting $\varphi_+ \to C_+, g_h \to g_c$:
\begin{align}
    \mathcal{L}_{C_+} =& \frac{2e^{4\sigma}}{r}\left[ |\dot C_+|^2 +|C_+|^2 \left(rM_5 g_c \right)^2
    -  rM_5 (\dot g_c + g_c\dot \sigma)  |C_+|^2 \right ] - \frac{3re^{4\sigma}}{8} \mathcal{R} |C_+|^2 \, .
\end{align}
Using the relation~\eqref{eq:gaugefixing} we find that the $|\dot C_+|^2$ term leads to quartic interactions (which we ignore), while the other terms combine to give
\begin{align}
    \mathcal{L}_{C_+} + \mathcal{L}_{\varphi_+} + \mathcal{L}_{\varphi_-}  =& - \frac{3 M_5^3 re^{4\sigma}}{8} \left( \mathcal{R} - \frac{16}{3} (M_5 g_c)^2 \right) - 2e^{4\sigma}M_5^4 (\dot g_c + g_c\dot \sigma) 
    \nonumber
    \\
    &-2r e^{4\sigma} \left[ r^{-2}\left( |\dot \varphi_+|^2 +  |\dot \varphi_-|\right)^2 + \left( |\varphi_+|^2 + |\varphi_-|^2\right) M_5^2 (g_h^2 +g_c^2) \right]  \, ,
    \\
    & + 2e^{4\sigma}M_5 \left[ (\dot g_h - \dot g_c + (g_h - g_c)\dot \sigma) |\varphi_+|^2 - (\dot g_h + \dot g_c + (g_h + g_c)\dot \sigma) |\varphi_-|^2  \right] \nonumber
\end{align}
We now want to add the terms from $\mathcal{L}_{\rm D}$, which can be written as
\begin{align}
    \mathcal{L}_{\rm D} 
    =& - \frac{M_5^3 re^{4\sigma}}{8} \left( \mathcal{R} - \frac{16}{3} (M_5 g_c)^2 \right)
    + \frac{4 g_c M_5^2 r}{3} e^{4\sigma} \left((g_c+g_h)|\varphi_-|^2 + (g_c - g_h) |\varphi_+|^2 \right) \, ,
    \nonumber
    \\
    & + 2 g_c M_5e^{4\sigma} \dot \sigma \left( M_5^3 + (g_c+g_h)|\varphi_-|^2 + (g_c - g_h) |\varphi_+|^2 \right) + \mathcal{O}(|\varphi|^4)  \, .
\end{align}

Putting everything together we find that the terms proportional to a single power of $\dot \sigma$ cancel and we are left with
\begin{align}
    \mathcal{L}_{C} + \mathcal{L}_{\varphi_+} + \mathcal{L}_{\varphi_-} + \mathcal{L}_{\rm D} =& - \frac{M_5^3 re^{4\sigma}}{2} \left( \mathcal{R} - 12 k^2 \right) 
    - 2r e^{4\sigma} \left[ r^{-2}\left( |\dot \varphi_+|^2 +  |\dot \varphi_-|\right)^2 + m_+^2 |\varphi_+|^2 + m_-^2 |\varphi_-|^2 \right] \, , \nonumber
   \\ 
   &+ 2e^{4\sigma}M_5 \left[ (\dot g_h - \dot g_c ) |\varphi_+|^2 - (\dot g_h + \dot g_c ) |\varphi_-|^2 - M_5^3 \dot g_c   \right] \, ,
   \label{eq:L_quad}
\end{align}
after identifying $12k^2 = 16(M_5 g_c)^2/3$. The bulk masses for the scalars are
\begin{align}
    m_\pm^2 =  \frac{M_5^2}{3} \left( 3 g_h^2 + g_c^2 \pm 2g_c g_h \right) \, .
\end{align}
The mass splittings for the bulk scalars arise due to the fact that the coupling of $\varphi_+$ and $\varphi_-$ to $V$ have opposite signs, which is a consequence of the orbifold projection and the gauging of the $U(1)$ generated by $\sigma_3$. The mass difference between the two fields is therefore a manifestation of the breaking of $\mathcal{N}=2$ SUSY by orbifolding. We note also that the fields $\varphi_\pm$ are not canonically normalized, but as we are concerned with their classical profiles this does not affect our results. 

The second line of equation~\eqref{eq:L_quad} leads to boundary terms after using the fact that the couplings are $\mathbb{Z}_2$-odd:
\begin{align}
    \dot g_{c/h} = 2\left[\delta(y) - \delta(y-\pi) \right] \widetilde g_{c/h} \, .
\end{align}
The boundary tensions we find are:
\begin{align}
    \mathcal{L} \big |_{\rm bdy} &= - \left[\delta(y) - \delta(y-\pi)\right] 8 M_5 e^{4\sigma} \left( g_c M_5^3 + (g_c + g_h) |\varphi_-|^2 - (g_h - g_c) |\varphi_+|^2 \right) \, ,
\end{align}
which was quoted in equation~\eqref{eq:tensions}.

\bibliographystyle{utphys}
\bibliography{draft.bib}{}

\end{document}